\documentclass[acmsmall, nonacm]{acmart}
\usepackage{microtype}
\usepackage{graphicx}
\usepackage{booktabs}
\usepackage{array}
\usepackage{amsmath}
\usepackage{enumitem}
\usepackage{xspace}
\usepackage{pifont}
\usepackage{xcolor}
\usepackage{tikz}
\usetikzlibrary{positioning,arrows.meta,calc}
\usepackage{adjustbox}
\usepackage{hyperref}
\usepackage{algorithm}
\usepackage{algorithmic}
\usepackage{listings}

\newcommand{\vpEtoEMeanBbhLow}{-4.6}

\newcommand{\vpEtoEMeanBbhMid}{-13.6}

\newcommand{\vpEtoEMeanCoqaLow}{-8.0}

\newcommand{\vpEtoEMeanGsmMid}{-36.8}

\IfFileExists{generated/capacity_macros.tex}{

\newcommand{\vpCapGsmRpsAbs}{11.3}

\newcommand{\vpCapBbhRps}{+7.4}

\newcommand{\vpCapBbhRpsAbs}{7.4}
\newcommand{\vpCapBbhTps}{+4.1}

\newcommand{\vpCapDoneArrivedMin}{0.50}
\newcommand{\vpCapDoneArrivedMax}{0.80}
}{}   
\IfFileExists{generated/headline_macros_abs.tex}{

\newcommand{\vpEtoEMeanBbhMidAbs}{13.6}

\newcommand{\vpMakespanCoqaLowAbs}{7.8}

\newcommand{\vpMakespanCoqaMidAbs}{7.0}

\newcommand{\vpTTFTMeanGsmMidAbs}{66.0}

\newcommand{\vpEtoEMeanGsmMidAbs}{36.8}

}{}   
\IfFileExists{generated/absolute_macros.tex}{

\newcommand{\vpAbsEtoEGsmMid}{26.9}

\newcommand{\vpAbsEtoEBbhMid}{23.9}

\newcommand{\vpAbsDrainShareCoqaLow}{61}

\newcommand{\vpAbsEtoECoqaMid}{13.1}

\newcommand{\vpAbsDrainShareCoqaHigh}{76}
}{}   
\IfFileExists{generated/engagement_macros.tex}{

}{}   
\IfFileExists{generated/sweep_vstar_macros.tex}{

\newcommand{\vpVstarTauMsLSixteen}{2.67}

\newcommand{\vpVstarPeakTbps}{1.94}
}{}   
\IfFileExists{generated/attribution_macros.tex}{\newcommand{\vpAttrRawVskN}{143}

\newcommand{\vpAttrFdBOneVskPct}{20.3}

\newcommand{\vpAttrFdBOneCtlPct}{2.0}

\newcommand{\vpAttrFdPopN}{500}

\newcommand{\vpAttrFdPopPct}{2.2}

\newcommand{\vpAttrFdPopCi}{1.3}

}{}   
\IfFileExists{generated/qwen_quality_macros.tex}{

}{}   
\IfFileExists{generated/quality_macros.tex}{\newcommand{\vpQGsmBminusA}{-3.26}

\newcommand{\vpQGsmDod}{-0.23}
\newcommand{\vpQGsmDodCi}{0.68}
\newcommand{\vpQGsmNdocs}{1,319}

\newcommand{\vpQCoqaDod}{+0.07}
\newcommand{\vpQCoqaDodCi}{0.12}
\newcommand{\vpQCoqaNdocs}{500}

\newcommand{\vpQBbhDod}{+0.45}
\newcommand{\vpQBbhDodCi}{0.55}
\newcommand{\vpQBbhNdocs}{6,511}

\newcommand{\vpQGsmDodStrict}{+0.15}
\newcommand{\vpQGsmDodStrictCi}{0.66}
\newcommand{\vpQGsmDodFlex}{-0.38}
\newcommand{\vpQGsmDodFlexCi}{0.74}
\newcommand{\vpQGsmBminusAStrict}{+5.31}

\newcommand{\vpQGsmBminusAFlex}{-5.53}

\newcommand{\vpQGsmArmAStrictUnparsed}{271}
\newcommand{\vpQGsmArmAStrictUnparsedPct}{20.5}
\newcommand{\vpQGsmArmAFlexFooled}{0}
\newcommand{\vpQGsmArmAFlexFooledPct}{0.0}
\newcommand{\vpQGsmArmBStrictUnparsed}{121}
\newcommand{\vpQGsmArmBStrictUnparsedPct}{9.2}
\newcommand{\vpQGsmArmBFlexFooled}{38}
\newcommand{\vpQGsmArmBFlexFooledPct}{2.9}
\newcommand{\vpQGsmArmCStrictUnparsed}{274}
\newcommand{\vpQGsmArmCStrictUnparsedPct}{20.8}
\newcommand{\vpQGsmArmCFlexFooled}{0}
\newcommand{\vpQGsmArmCFlexFooledPct}{0.0}
\newcommand{\vpQGsmArmDStrictUnparsed}{121}
\newcommand{\vpQGsmArmDStrictUnparsedPct}{9.2}
\newcommand{\vpQGsmArmDFlexFooled}{39}
\newcommand{\vpQGsmArmDFlexFooledPct}{3.0}

}{}
\IfFileExists{generated/natural_lengths_macros.tex}{

\newcommand{\vpNatLenUpRunawayPctCoqaLow}{2.0}

\newcommand{\vpNatLenUpRunawayTokPctCoqaLow}{94}

\newcommand{\vpNatLenUpShortPctCoqaMid}{43}

}{}
\IfFileExists{generated/latency_map_macros.tex}{\newcommand{\vpMapSpreadBbhLow}{0.9}

\newcommand{\vpMapSpreadBbhMid}{3.6}

\newcommand{\vpMapSpreadCoqaLow}{11.1}

\newcommand{\vpMapSpreadCoqaHigh}{7.8}

\newcommand{\vpMapSpreadGsmLow}{0.7}

\newcommand{\vpMapSpreadGsmMid}{1.3}

\newcommand{\vpMapSpreadGsmHigh}{8.0}

}{}
\IfFileExists{generated/sweep_macros.tex}{

\newcommand{\vpSweepEtoEMeanRFiftyDFifty}{-3.6}

\newcommand{\vpSweepEtoEMeanBest}{-49.1}

}{}
\IfFileExists{generated/sweep_prediction_macros.tex}{

\newcommand{\vpPredUpstreamOcc}{411}

}{}
\IfFileExists{generated/sweep_macros_v2.tex}{

}{}
\IfFileExists{generated/sweep_macros_ungated.tex}{

\newcommand{\vpSweepUngEtoEMeanBest}{-62.9}

\newcommand{\vpSweepUngEtoEMeanWorst}{+23.8}

}{}
\IfFileExists{generated/sweep_ttft_macros.tex}{

\newcommand{\vpSweepTtftMeanRFiftyDFifty}{-33.6}

\newcommand{\vpSweepTtftMeanRSeventyfiveDFifty}{-61.2}

\newcommand{\vpSweepTtftMeanBest}{-78.2}

\newcommand{\vpSweepTtftMeanWorst}{+44.2}

}{}   
\IfFileExists{generated/sweep_ttft_macros_ungated.tex}{

}{}
\IfFileExists{generated/sweep_ttft_macros_v2.tex}{

}{}
\IfFileExists{generated/natural_lane_macros.tex}{

\newcommand{\vpNatGsmMidOutTok}{+185}

\newcommand{\vpNatGsmMidCapUp}{3}
\newcommand{\vpNatGsmMidCapVs}{99}
\newcommand{\vpNatGsmMidOutTokUp}{116}
\newcommand{\vpNatGsmMidOutTokVs}{330}

\newcommand{\vpNatBbhLowOutTokAbs}{53}

\newcommand{\vpNatBbhMidOutTokAbs}{59}

\newcommand{\vpNatBbhMidOutTokUp}{255}

\newcommand{\vpNatCoqaMidOutTok}{-24}

\newcommand{\vpNatCoqaMidEtoE}{-26}

\newcommand{\vpNatCoqaMidOutTokUp}{164}

\newcommand{\vpNatCoqaHighOutTok}{-3}

\newcommand{\vpNatCoqaHighEtoE}{-4}

}{}
\IfFileExists{generated/loaded_quality_macros.tex}{

\newcommand{\vpLqGsmLowDeltaUp}{-0.38}
\newcommand{\vpLqGsmLowDeltaUpCi}{1.39}

\newcommand{\vpLqGsmMidDelta}{-0.08}
\newcommand{\vpLqGsmMidDeltaCi}{1.46}

\newcommand{\vpLqCoqaMidDelta}{-0.05}
\newcommand{\vpLqCoqaMidDeltaCi}{1.76}

\newcommand{\vpLqBbhMidDelta}{-0.22}
\newcommand{\vpLqBbhMidDeltaCi}{0.38}

}{}
\IfFileExists{generated/client_equivalence_macros.tex}{

}{}
\IfFileExists{generated/loaded_shares_macros.tex}{

\newcommand{\vpLqGsmMidHybRoutedShare}{89.5}

\newcommand{\vpLqGsmHighHybRoutedShare}{98.7}

\newcommand{\vpLqBbhMidHybRoutedShare}{99.0}

}{}
\IfFileExists{generated/f1_macros.tex}{
\newcommand{\vpFOneDropBsOne}{21.0}
\newcommand{\vpFOneDropBsEight}{14.6}

}{}
\IfFileExists{generated/cache_share_macros.tex}{\newcommand{\vpCacheGsm}{1.7}
\newcommand{\vpCacheGsmVs}{1.4}
\newcommand{\vpCacheBbh}{91.0}
\newcommand{\vpCacheBbhVs}{90.2}
\newcommand{\vpCacheCoqa}{1.1}
\newcommand{\vpCacheCoqaVs}{1.1}
\newcommand{\vpCacheArmDiffMax}{0.8}
}{}
\IfFileExists{generated/qwen_serving_macros_v16.tex}{

\newcommand{\vpQwenFourEtoEMeanQwenFourMid}{+2.7}
\newcommand{\vpQwenFourEtoEMeanQwenFourMidCi}{2.8}

\newcommand{\vpQwenEightEtoEMeanQwenEightMidCi}{0.3}

}{}   
\IfFileExists{generated/qwen_blockb_macros.tex}{

\newcommand{\vpLqQwenFourDelta}{+1.52}
\newcommand{\vpLqQwenFourDeltaCi}{2.49}

\newcommand{\vpLqQwenFourHybRoutedShare}{86.5}

\newcommand{\vpLqQwenEightDelta}{+4.25}
\newcommand{\vpLqQwenEightDeltaCi}{2.68}

\newcommand{\vpLqQwenEightHybRoutedShare}{99.3}

}{}   
\IfFileExists{generated/attested_skip_macros.tex}{

}{}
\IfFileExists{generated/served_skip_macros.tex}{
\newcommand{\vpLlamaKneeSDecode}{0.504}

\newcommand{\vpQwenFourKneeSDecode}{0.119}

\newcommand{\vpQwenEightKneeSDecode}{0.416}

\newcommand{\vpQwenEightAltKneeSDecode}{0.350}

\newcommand{\vpQwenEightOldKneeSDecode}{0.345}

}{}   
\IfFileExists{generated/probe_skip_macros.tex}{
\newcommand{\vpProbeSkipQwenEight}{0.520}
\newcommand{\vpProbeSkipQwenEightAlt}{0.485}
\newcommand{\vpProbeSkipQwenEightOld}{0.445}
}{}   
\IfFileExists{generated/ladder_macros.tex}{\newcommand{\vpLadderGsmKnee}{13}

\newcommand{\vpLadderBbhKnee}{34}

\newcommand{\vpLadderCoqaKnee}{25}

}{}   
\IfFileExists{generated/ladder_h100_macros.tex}{\newcommand{\vpLadderHundredGsmKnee}{14}

}{}   
\IfFileExists{generated/ladder_a6000_macros.tex}{\newcommand{\vpLadderAsixGsmKnee}{6}

}{}   
\IfFileExists{generated/a6000_macros.tex}{

\newcommand{\vpAsixEtoEMeanGsmMidCi}{4.2}

}{}   
\IfFileExists{generated/transfer_macros_abs.tex}{

\newcommand{\vpHoneEtoEMeanGsmMidAbs}{39.4}

\newcommand{\vpAsixEtoEMeanGsmMidAbs}{30.0}

\newcommand{\vpQwenEightEtoEMeanQwenEightMidAbs}{6.6}

}{}   
\IfFileExists{generated/tile_ratio_macros.tex}{
\newcommand{\vpTileRatioMedAHundred}{1.18}
\newcommand{\vpTileRatioMinAHundred}{0.82}
\newcommand{\vpTileRatioMaxAHundred}{1.79}

\newcommand{\vpTileRatioMedAsix}{1.13}
\newcommand{\vpTileRatioMinAsix}{0.77}
\newcommand{\vpTileRatioMaxAsix}{1.53}

}{}   
\IfFileExists{generated/hardware_bands_macros.tex}{

\newcommand{\vpBandTauMs}{2.67}
\newcommand{\vpBandSkipRatio}{0.5}

}{}
\IfFileExists{generated/pareto_macros.tex}{

}{}
\IfFileExists{generated/pareto_models_macros.tex}{
\newcommand{\vpPMLlamaGsmLat}{-36.8}
\newcommand{\vpPMLlamaGsmLatCi}{2.0}
\newcommand{\vpPMLlamaGsmQ}{-3.64}
\newcommand{\vpPMLlamaGsmQCi}{2.22}
\newcommand{\vpPMLlamaGsmLatUp}{26.9}
\newcommand{\vpPMLlamaGsmLatHyb}{17.0}
\newcommand{\vpPMLlamaGsmQUp}{77.6}
\newcommand{\vpPMLlamaGsmQHyb}{74.0}
\newcommand{\vpPMQwenFourLat}{+2.7}
\newcommand{\vpPMQwenFourLatCi}{2.8}
\newcommand{\vpPMQwenFourQ}{-2.96}
\newcommand{\vpPMQwenFourQCi}{2.15}
\newcommand{\vpPMQwenFourLatUp}{26.7}
\newcommand{\vpPMQwenFourLatHyb}{27.4}
\newcommand{\vpPMQwenFourQUp}{80.7}
\newcommand{\vpPMQwenFourQHyb}{77.7}
\newcommand{\vpPMQwenEightLat}{-6.6}
\newcommand{\vpPMQwenEightLatCi}{0.3}
\newcommand{\vpPMQwenEightQ}{-4.47}
\newcommand{\vpPMQwenEightQCi}{2.33}
\newcommand{\vpPMQwenEightLatUp}{71.6}
\newcommand{\vpPMQwenEightLatHyb}{66.8}
\newcommand{\vpPMQwenEightQUp}{80.4}
\newcommand{\vpPMQwenEightQHyb}{75.9}
}{}   
\IfFileExists{generated/graph_coverage_macros.tex}{

\newcommand{\vpCovBbhMidUpAboveTwoFiftySix}{39}

\newcommand{\vpCovBbhHighUpAboveOneKtwentyFour}{23}

\newcommand{\vpCovBbhMidVsAboveTwoFiftySix}{40}

\newcommand{\vpCovBbhHighVsAboveOneKtwentyFour}{23}

\newcommand{\vpCovCoqaMidUpAboveTwoFiftySix}{5}

\newcommand{\vpCovCoqaMidVsAboveTwoFiftySix}{6}

\newcommand{\vpCovGsmMidUpAboveTwoFiftySix}{57}

\newcommand{\vpCovGsmMidVsAboveTwoFiftySix}{20}

}{}   
\IfFileExists{generated/suite_macros.tex}{\newcommand{\vpSuiteGsmRequests}{3,600}

\newcommand{\vpSuiteBbhRequests}{4,000}

\newcommand{\vpSuiteCoqaRequests}{4,000}

\newcommand{\vpSuiteRemainingCtxMin}{6,171}
\newcommand{\vpSuiteRemainingCtxMax}{7,988}
}{}   
\IfFileExists{generated/h100_macros.tex}{

\newcommand{\vpHoneEtoEMeanGsmMid}{-39.4}
\newcommand{\vpHoneEtoEMeanGsmMidCi}{0.2}

}{}   
\IfFileExists{generated/ablation_macros.tex}{

\newcommand{\vpAblVpreBinarycohortEtoEMid}{-35.6}

\newcommand{\vpAblVskipperNoupgradeEtoEMid}{-32.6}

}{}
\IfFileExists{generated/ablation_macros_bbh_cot.tex}{

\newcommand{\vpAblBbhVdecFdEtoEMid}{-16.0}

}{}
\IfFileExists{generated/ablation_macros_coqa.tex}{

\newcommand{\vpAblCoqaIntegratedAlwaysskipEtoEMid}{+5.5}

\newcommand{\vpAblCoqaIntegratedAlwaysskipEtoEMidVsHeadAbs}{9.5}

}{}
\newif\ifvpUpstreamBanks
\IfFileExists{generated/bank_policy.tex}{\vpUpstreamBankstrue
}{\vpUpstreamBankstrue}

\providecommand{\vpQGsmDod}{\textbf{[pending]}}
\providecommand{\vpQBbhDod}{\textbf{[pending]}}
\providecommand{\vpMapSpreadGsmMid}{\textbf{[pending]}}\providecommand{\vpMapSpreadBbhMid}{\textbf{[pending]}}

\providecommand{\vpSweepEtoEMeanBest}{\textbf{[pending]}}


\newcolumntype{Z}{>{\setbox0=\hbox\bgroup}l<{\egroup}@{}}
\lstdefinestyle{vpy}{
  language=Python, basicstyle=\ttfamily\scriptsize, keywordstyle=\bfseries,
  commentstyle=\itshape, showstringspaces=false, columns=fullflexible,
  breaklines=true, frame=none, xleftmargin=0pt, aboveskip=4pt, belowskip=4pt,
  morekeywords={frozenset}
}

\definecolor{accent}{HTML}{193A5A}
\definecolor{controlfill}{HTML}{EEF3F8}
\definecolor{runtimefill}{HTML}{FFF4E8}
\definecolor{statefill}{HTML}{EEF7F0}
\definecolor{neutralfill}{HTML}{F5F6F7}
\definecolor{linecolor}{HTML}{52616F}
\definecolor{winc}{HTML}{1B7837}
\definecolor{lossc}{HTML}{B2182B}
\definecolor{vpaccent}{HTML}{193A5A}
\definecolor{vplinecolor}{HTML}{52616F}
\newcommand{\vskipper}{\textsc{vSkipper}\xspace}
\newcommand{\sys}{\textsc{vSkipper}}

\setlist[enumerate]{leftmargin=1.35em,itemsep=1pt,topsep=2pt,parsep=0pt}
\setlist[itemize]{leftmargin=1.2em,itemsep=1pt,topsep=2pt,parsep=0pt}
\title[\vskipper]{\vskipper: Translating Dynamic Layer Skipping\\
into LLM Serving Gains}
\author{Wei Da}
\authornote{Corresponding author.}
\affiliation{%
  \institution{The University of Cambridge}
  \city{Cambridgeshire}
  \country{United Kingdom}}
\email{wd312@cam.ac.uk}

\author{Yavuz Ferhatosmanoglu}
\affiliation{%
  \institution{The University of Cambridge}
  \city{Cambridgeshire}
  \country{United Kingdom}}
\email{yff23@cantab.ac.uk}

\author{Evangelia Kalyvianaki}
\affiliation{%
  \institution{The University of Cambridge}
  \city{Cambridgeshire}
  \country{United Kingdom}}
\email{ek264@cam.ac.uk}
\providecommand{\vpLadderAsixGsmKnee}{6}
\begin{document}

\begin{abstract}
Dynamic layer skipping reduces LLM computation by allowing each token to execute only a subset of the model's layers. However, existing skippers rely on specialized generation loops and do not integrate with modern serving engines. As a result, fewer executed layers do not necessarily translate into lower serving latency: FlexiDepth skips $8$ of Llama-3-8B's $32$ layers on average, yet its standard generation loop decodes $\vpFOneDropBsEight$--$\vpFOneDropBsOne\%$ more slowly than the base model. We present \sys{}, a virtualization layer that makes dynamic layer skippers pluggable in serving engines while preserving continuous batching, fixed-shape batches, paged KV caching, and captured decode graphs. At each routed layer, \sys{} groups tokens by the skipper's decision and uses routed execution only when predicted to be profitable. We implement \sys{} in SGLang and evaluate the released FlexiDepth checkpoint against upstream SGLang under identical prompts, arrivals, output lengths, and launch settings. At the knee of upstream's load curve, \sys{} reduces mean end-to-end latency by $\vpEtoEMeanGsmMidAbs\%$ on GSM8K and $\vpEtoEMeanBbhMidAbs\%$ on BBH. Under saturation, it increases request throughput by $\vpCapGsmRpsAbs\%$ and $\vpCapBbhRpsAbs\%$. Serving adds no statistically resolved quality loss beyond the checkpoint's own. Across synthetic skip policies, two Qwen3 skippers, and three GPUs, we demonstrate reuse without workload-specific tuning. To our knowledge, \sys{} is the first system to realize serving-efficiency gains from per-token interior layer skipping within a modern LLM serving engine. The code is open-sourced as an SGLang fork at \textcolor{blue}{\url{https://github.com/AKafakA/sglang-vskipper/tree/vskipper-ref}}
\end{abstract}

\maketitle

\section{Introduction}
\label{sec:intro}

Not every token requires the full depth of an LLM. Dynamic layer-skipping methods exploit this observation by choosing, for each token, which layers to execute~\citep{luo2025flexidepth, he2025adaskip, yang2025dash}. FlexiDepth~\citep{luo2025flexidepth}, for example, adds lightweight per-token routers to a pretrained model and skips $20$--$30\%$ of its layers. Such methods reduce model computation without replacing the underlying transformer backbone. However, translating saved computation into batched serving gains is a separate execution problem.

Modern LLM serving engines rely on regular batched execution, continuous batching, paged KV-cache management, and captured GPU graphs~\citep{yu2022orca,kwon2023pagedattention,zheng2024sglang}. These optimizations largely assume that tokens in a batch traverse every layer.
Dynamic skipping breaks this regularity: at a given layer, some tokens execute the full computation while others bypass it. 
Released skippers therefore commonly use model-specific research execution
paths and report FLOP savings~\citep{luo2025flexidepth}, rather than gains in modern serving engines. The missing piece is an interface between layer skipping and the engine's batching, cache management, and execution.

We present \sys{}, a serving virtualization layer for
per-token interior layer skipping. \sys{} makes a dynamic skipper a pluggable component of the serving engine through a reusable interface. The skipping policy decides where each token skips, while \sys{} preserves the engine's batching, cache, and graph interfaces. We realize mixed-depth execution through a RUN/\emph{Project-Only} interface. At each routed layer, \sys{} packs rows sharing a decision into cohorts and scatters their outputs back to the original batch positions. Project-Only rows bypass the layer's attention and FFN while still writing the KV state required by future tokens' attention computation. Since routing introduces overhead, \sys{} performs \textit{mode switching} only when its estimated savings exceed the cost.

We implement \sys{} in SGLang and evaluate it against upstream SGLang using identical launch settings,
prompts, arrivals and per-request output lengths. At the knee of upstream's load curve, mean end-to-end latency falls by
$\vpEtoEMeanGsmMidAbs\%$ on GSM8K and $\vpEtoEMeanBbhMidAbs\%$ on BBH. Under saturation, request throughput rises by $\vpCapGsmRpsAbs\%$
and $\vpCapBbhRpsAbs\%$. Serving introduces no statistically resolved quality loss beyond that of the checkpoint, and under load we find no resolved quality difference from routing every pass at the knee. To our knowledge, \sys{} is the first serving system to realize
serving-efficiency gains from per-token interior whole-layer
skipping within a modern LLM serving engine. 

Our contributions are:

\begin{enumerate}
\item \textbf{A virtualization layer for skippers.} We introduce a whole-layer RUN/Project-Only interface that makes per-token skip policies pluggable components of a modern serving engine without redesigning its scheduler or memory manager (\S\ref{sec:design}).

\item \textbf{Mixed-depth execution.} We develop route-guided cohort execution for heterogeneous per-token paths within captured graphs (\S\ref{sec:design-production}),
establish the invariants required for correct dynamic-depth execution  (\S\ref{sec:design-invariants}),
and introduce profitability-aware mode switching
(\S\ref{sec:design-engagement}).

\item \textbf{A controlled evaluation.} We compare \sys{} with upstream SGLang under matched work and launch settings, and separate checkpoint quality from serving effects. We then demonstrate how the same implementation generalizes beyond FlexiDepth across nine synthetic skip policies, two Qwen3 skippers in addition to Llama-3-8B, three workloads, and A100, H100, and RTX A6000 GPUs, without workload-specific tuning (\S\ref{sec:evaluation}).
\end{enumerate}

\section{Background and Motivation}
\label{sec:motivation}

\noindent\textbf{Serving engines assume fixed-depth execution.} Continuous-batching engines such as Orca~\citep{yu2022orca}, vLLM~\citep{kwon2023pagedattention}, and SGLang~\citep{zheng2024sglang} organize kernels, metadata, page tables, and captured graphs around the assumption that every row in a batch traverses every layer. Dynamic layer skipping breaks this assumption because tokens follow different layer paths. 

\begin{figure}[t]
  \centering
  \includegraphics[width=0.64\linewidth]{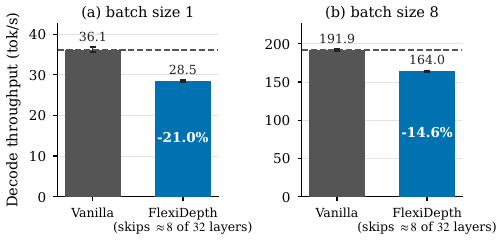}
  \caption{Saved FLOPs are not saved time: decode throughput of the
  FlexiDepth checkpoint against its base model in a standard generation
loop, with fixed batches of 1 and 8 requests each generating 256 tokens
(mean of three runs; whiskers show the range).}
  \label{fig:f1-motivation}
\end{figure}

\noindent\textbf{Saved FLOPs are not saved time.}
FlexiDepth reports that its implementation ``does not lead to improved throughput on the existing GPU hardware''~\citep{luo2025flexidepth}. We reproduce this gap using the public checkpoint over $100$ prompts with $256$ greedy output tokens. Despite skipping $8$ of $32$ layers on average, decode throughput falls by $\vpFOneDropBsOne\%$ at batch size~1 and $\vpFOneDropBsEight\%$ at batch size~8 (Figure~\ref{fig:f1-motivation}). At batch size~1, routing overhead offsets the saved work. At batch size~8, the batch still traverses the union of its tokens' required layers, while splitting it sacrifices batch width.

\noindent\textbf{What a serving abstraction must provide.} Turning dynamic depth into serving gains requires preserving batch efficiency, maintaining correct per-layer state across divergent paths, and routing only when the savings exceed the overhead. For autoregressive attention, even a row that skips a layer's main computation must still produce the KV state required by future tokens. The benefit also depends on phase: skipping reduces attention memory traffic during memory-bound decode and avoids arithmetic during compute-bound prefill.

\section{Design and Implementation}
\label{sec:design}

\begin{figure*}[t]
  \centering
  \includegraphics[width=0.99\linewidth]{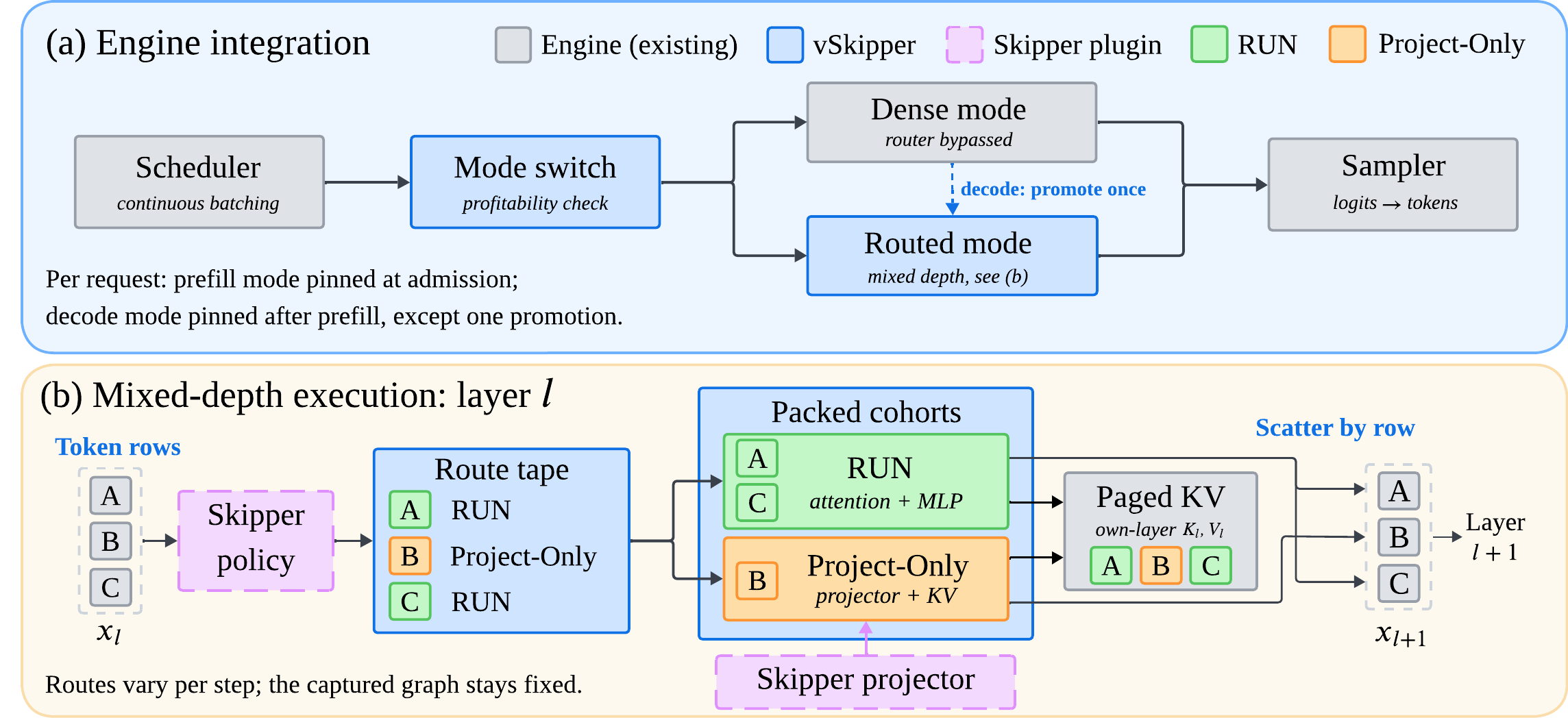}
  \caption{Overview of \sys{}. (a) The mode switch selects dense or routed execution per request and phase based on predicted profitability (\S\ref{sec:design-engagement}). (b) Routed layers record skipper decisions, pack rows into RUN and Project-Only cohorts, and scatter outputs back to the batch (\S\ref{sec:design-production}). Project-Only rows still write their own-layer KV (\S\ref{sec:design-invariants}).}
  \label{fig:arch-overview}
\end{figure*}

\sys{} separates the skipping policy from execution and engine state
(Figure~\ref{fig:arch-overview}). A skipper implements the \emph{skipper interface}: it supplies per-token RUN or Project-Only
decisions and the projector applied to skipped rows. We realize these
decisions through \emph{mixed-depth execution}, which turns them into
route data, packed cohorts, and KV-complete cache state
(\S\ref{sec:design-production}). Two invariants keep this execution
correct under dynamic depth (\S\ref{sec:design-invariants}), and the
\emph{mode switch} decides when routing is worth its cost
(\S\ref{sec:design-engagement}). The engine remains compatible with
its serving optimizations, including continuous batching, admission
control, paged KV management and prefix reuse
(\S\ref{sec:implementation}).

\emph{Vocabulary.} A \emph{row} holds one token's hidden state, one per sequence in decode and several per sequence in prefill. Each request uses \emph{dense} or \emph{routed} mode, chosen separately for prefill and decode. Dense mode is the unmodified engine's forward pass and runs every layer without consulting the router. Routed mode takes RUN or Project-Only per row at each routed layer. If a batch mixes modes, it uses the routed path with dense rows taking RUN. $V$ denotes a batch's resident KV tokens.

\subsection{Route-Guided Cohort Execution}
\label{sec:design-production}

Mixed-depth execution runs inside the engine's existing forward passes.
Decode steps run as in a standard engine, replaying a single captured graph over the
batch. Inside the graph, as Figure~\ref{fig:arch-overview} shows, rows in the same batch can take different actions chosen by the policy from the current rows, and the RUN and Project-Only rows compute their updates separately before scattering them back to their rows.
The next routed layer evaluates the updated states and may route differently.

\noindent\textbf{Route tape.} A layer's policy decisions are written 
into a device-resident \emph{route tape}, a per-row mask for each routed layer. Since the captured
graph consumes the tape as data, routes can change from step to step
without host synchronization or graph recapture.

\noindent\textbf{Project-Only path.} A skipped token cannot simply
bypass a layer: future tokens attending to it expect that
layer's KV to exist (Invariant~1, \S\ref{sec:design-invariants}).
Skipped layers are therefore traversed cheaply, with the skipper's own skip semantics. 
A skipped row writes only the layer's key and value projections to the KV cache, 
dropping its attention output and running the plugged projector instead of the MLP. 
For FlexiDepth, the projector is the
checkpoint's adapter, mixed by the router's weight $w \in [0,1]$.
The skip avoids the row's attention KV read and, when compute-bound, its bypassed arithmetic.

\noindent\textbf{Packed cohorts.} From the tape, each routed layer
derives its two groups of rows, the \emph{cohorts}, and their sizes
as device scalars. We treat each cohort as one contiguous operand,
folding the row gather into the operand load and the weighted
scatter of results into the epilogue, so the cohort GEMMs are
bounded by cohort size rather than batch size. Attention instead
keeps its launch shape and masks skipped rows' KV reads, leaving
the graph capture-safe (Algorithm~\ref{alg:compaction},
Appendix~\ref{app:compaction}).

\begin{algorithm}[t]
\caption{One routed layer of a decode graph captured at batch size
$C$. The cohort counts $n_r$, $n_p$ and router weights $w$ live on the device, so the layer
runs without host synchronization.}
\label{alg:compaction}
\begin{algorithmic}[1]
\REQUIRE the layer's route tape, hidden states $X$, residual $Y$, layer and projector weights, tile table
\ENSURE $Y$ updated for all rows and the layer's KV cache complete
\STATE \textbf{Maps.} From the route tape build the index maps
$\mathrm{run}[0{:}n_r]$ and $\mathrm{proj}[0{:}n_p]$, mapping each
cohort position to its batch row, and the counts $n_r$, $n_p$ as
device scalars.
\STATE \textbf{Attention.}  Compute and write every valid row's own-layer KV through the layer's normal projection and cache-write path. The decode kernel takes the RUN mask: a
Project-Only row reads no KV at this layer and emits zero.

\STATE \textbf{Tiles.} For each of the four count-bounded GEMMs
(gate--up and down, per cohort) select the tile configuration from the device's tuned tile table with captured batch size $C$.

\STATE \textbf{RUN cohort.} $G \gets \mathrm{GEMM}_{n_r}(X[\mathrm{run}],
W_{\mathrm{gate,up}})$ with the gather folded into the operand load;
$A \gets \mathrm{silu}(G_{\mathrm{gate}}) \odot G_{\mathrm{up}}$ over $n_r$
rows; $Y[\mathrm{run}[m]] \mathrel{+}= w\,\mathrm{GEMM}_{n_r}(A,
W_{\mathrm{down}})[m]$ with the scatter and the router weight $w$
folded into the epilogue.
\STATE \textbf{Project-Only cohort.} The same three steps over
$\mathrm{proj}$ and $n_p$, with the projector's weights and route
weight $1-w$.

\STATE \textbf{Grid exit.} Every GEMM launches a static grid sized
for $C$ rows, and a tile starting at or beyond the device-side cohort count exits immediately, so one captured graph serves every cohort size.
\end{algorithmic}
\end{algorithm}

\noindent\textbf{Captured execution.}  We capture both modes at
fixed batch sizes, up to a device-derived maximum (e.g.,
$1{,}024$ rows on an A100), so the mode switch always finds a captured graph. Within the routed graph, each routed layer runs the
mixed-cohort path of Algorithm~\ref{alg:compaction}, and a
device-side conditional CUDA-graph node adds an all-RUN fast path
for the frequent steps where no row skips the layer.

\noindent\textbf{Prefill.} The same interface and cohorts govern prefill, with one row per prompt token and its own switching rule
(\S\ref{sec:design-engagement}). Its variable-size inputs keep
prefill outside capture, so cohorts use standard cuBLAS GEMMs with
the same gather and scatter (Appendix~\ref{app:compaction}). In
this compute-bound phase, a skipped row saves arithmetic directly
and writes each layer's KV as in decode.

\subsection{Invariants for Correct Mixed-Depth Execution}
\label{sec:design-invariants}

\sys{} enforces two invariants that preserve correctness when depth
varies per token.

\noindent\textbf{Invariant 1: every layer writes its own KV.}
\emph{Every cached token has keys and values computed by each
layer's own projection weights at all $L$ layers, including the
skipped layers.} Batched attention, paged KV
caches~\citep{kwon2023pagedattention} and prefix reuse all assume
that position $t$ of layer $\ell$'s cache holds the outputs of that layer's own key and value projections, rather than copied or missing state. Early-exit methods can violate this
and must recompute, keep per-depth
caches~\citep{elhoushi2024layerskip}, or copy state across
depths~\citep{liu2025drex}. The Project-Only path preserves it.

\noindent\textbf{Invariant 2: row metadata is never stale.}
\emph{Any structure that maps rows to memory is recomputed from the
current row set wherever that set can change.} This
covers attention metadata, KV page mappings and every gather or
scatter map derived from routes. Under dynamic depth these maps
change per layer and step, and reusing a stale map can make
requests silently read other requests' KV pages. \sys{} therefore
derives every route-dependent map on the device from the current
tape.

\subsection{Profitability-Aware Mode Switching}
\label{sec:design-engagement}

Routed mode has a fixed per-step cost, so it is \emph{profitable} only when its savings exceed that cost. The mode switch is a host-side profitability check that selects dense or routed mode separately for prefill and decode. Prefill mode is fixed at admission, while decode mode is set after prefill and may be promoted once from dense to routed if load crosses the enter threshold, after which it never reverts. Because this promotion can change computation mid-generation, we evaluate the resulting \emph{served hybrid} directly (\S\ref{sec:eval-q2}). The \emph{always-route} setting disables the switch.

\noindent\textbf{Decode.} Below the device's \emph{ridge point}, where a kernel becomes compute-bound in the roofline model~\citep{williams2009roofline}, decode is limited by weight bandwidth, and a skipped row saves only its KV read. Let $L_r$ be the number of routed layers, $s$ the fraction of decisions
projecting, $b$ the bytes of KV per token and layer, and $\tau$
routed mode's fixed per-step cost. Routing is profitable when
\[
  V \;>\; V^* = \frac{\tau \cdot \mathrm{BW}}{s\,L_r\,b}
  \;\approx\; 158\text{k resident tokens on an A100}
\]
($\tau = \vpVstarTauMsLSixteen$\,ms, $s = 0.5$, $L_r = 16$,
$b = 4$\,KB; Appendix~\ref{app:roofline}). The served \emph{switching thresholds} form a hysteresis band: routing is enabled at $V \geq 200$k and disabled for new decode requests at $V \leq 160$k. Within this range, the current switching state is retained.
The rule reproduces the served exit threshold within $2\%$. Every other deployment's thresholds follow from the rule with its device and
model constants (Appendix~\ref{app:roofline}).

\noindent\textbf{Prefill.} Prefill uses routed mode for batches with at least $1{,}536$ prompt tokens and an estimated Project-Only share above $0.35$, jointly sufficient to cover the fixed cost. The share is probed every $64$ admitted batches. These constants are fixed across workloads (Appendix~\ref{app:roofline}).

\subsection{Implementation}
\label{sec:implementation}

We implement \sys{} as a fork of SGLang~\citep{zheng2024sglang},
leaving unchanged its multiprocess architecture, admission control
and memory management. The scheduler additionally records each request's mode, which also keys
prefix reuse. Routed-layer hooks support the 
Llama-3 and Qwen3 families. The main skipper is the released FlexiDepth checkpoint for Llama-3-8B~\citep{luo2025flexidepth}, 
routing the last $16$ of $32$
layers, and we train and serve two Qwen3 skippers (Appendix~\ref{app:qwen}).

Under the skipper interface, we implement policies of two kinds: a
trained gate (FlexiDepth and the Qwen3 skippers) and a seeded hash
of request and token (RandomSkip). A new policy can reuse an
existing projector and its execution path; only a new projector 
requires its own loading and execution code and checkpoint tensors.
At startup, the runtime rejects any policy whose actions fall outside 
the whole-layer \{RUN, Project-Only\} set. 
Appendix~\ref{app:plugin} illustrates
the interface with a static-depth policy.

\section{Evaluation}
\label{sec:evaluation}

We present the experimental setup
(\S\ref{sec:methodology}) and then evaluate \sys{} on three
questions. \textbf{(Q1)} Does it turn a skipper's serving deficit
into a gain at matched work (\S\ref{sec:eval-q1})? \textbf{(Q2)}
Does it add any quality loss beyond the skipper's own
(\S\ref{sec:eval-q2})? \textbf{(Q3)} Is the abstraction general
across skippers, models, workloads and hardware
(\S\ref{sec:eval-q3})?

\subsection{Experimental Setup}
\label{sec:methodology}

\noindent\textbf{Arms.} We define an \emph{arm} as one measured
configuration and a \emph{cell} as one (workload, offered rate)
pair. The baseline arm is upstream SGLang, the code our fork
derives from, served from its own checkout. The \sys{} arms are the
served \emph{hybrid} and always-route settings
(\S\ref{sec:design-engagement}). All arms share one launch profile
(fp16 for Llama, bf16 for Qwen, static memory fraction $0.8$,
Triton attention), capture decode graphs at the same batch sizes,
and keep SGLang's radix prefix cache enabled, with \sys{} maintaining mode-consistent prefix reuse (Appendix~\ref{app:absolutes}). We check every arm's reported
execution settings for undeclared differences
(Appendix~\ref{app:gates}).

\noindent\textbf{Load.} We set load through the offered arrival
rate rather than admission control: seeded Poisson arrivals, no
client-side concurrency cap, and every request served to
completion. For each workload we measure $Q^*$, the last offered
rate at which the baseline's output token rate still grows under
its own natural generation (Appendix~\ref{app:repro}). We call
$0.95\times Q^*$ the \emph{knee}, $0.75\times Q^*$ \emph{below the knee}
and $1.25\times Q^*$ \emph{overload}, evaluating arms at those
absolute rates.

\noindent\textbf{Workloads.} We evaluate on three standard tasks
with complementary serving profiles (Table~\ref{tab:workloads}):
GSM8K mathematical reasoning~\citep{cobbe2021gsm8k}, pairing long
few-shot prompts with short outputs; BBH (BIG-Bench
Hard)~\citep{suzgun2023bbh}, whose chain-of-thought generations
result in long outputs and whose shared few-shot prefixes are largely
served from the prefix cache; and CoQA conversational
QA~\citep{reddy2019coqa}, whose long passages and very short
answers ($\vpNatLenUpShortPctCoqaMid\%$ end within five tokens)
concentrate decode work in a few long requests. We serve each workload as a
frozen suite of token-id prompts, padding the GSM8K and CoQA load
with training-split prompts and scoring only their held-out documents.

\begin{table}[t]
  \centering
  \caption{The three workloads: serving protocol, requests per
  performance cell, and documents per quality cell. Output tokens
  and E2E are upstream's mean values at its knee.}
  \label{tab:workloads}
  \scriptsize
  \begin{tabular}{llrrrrr}
  \toprule
  Workload & protocol & requests & documents & output tokens & $Q^*$ (req/s) & upstream E2E (s) \\
  \midrule
  GSM8K & 5-shot, chat template & $\vpSuiteGsmRequests$ & $\vpQGsmNdocs$ & $\vpNatGsmMidOutTokUp$ & $\vpLadderGsmKnee$ & $\vpAbsEtoEGsmMid$ \\
  BBH & 3-shot chain-of-thought & $\vpSuiteBbhRequests$ & $\vpQBbhNdocs$ & $\vpNatBbhMidOutTokUp$ & $\vpLadderBbhKnee$ & $\vpAbsEtoEBbhMid$ \\
  CoQA & raw completion & $\vpSuiteCoqaRequests$ & $\vpQCoqaNdocs$ & $\vpNatCoqaMidOutTokUp$ & $\vpLadderCoqaKnee$ & $\vpAbsEtoECoqaMid$ \\
  \bottomrule
  \end{tabular}
\end{table}

\noindent\textbf{Matched work.} The arms generate the same number of
output tokens for each request and differ only in how they execute that
work. Under natural
stopping, the checkpoint's output lengths differ from the base model's
(Appendix~\ref{app:banks}), which would confound execution speed with generation behavior. We therefore record each request's output length
from the baseline's natural generation at the cell's rate and replay
it in both arms with \texttt{ignore\_eos}. Prompts, arrival schedule, output lengths
and launch profile are identical across arms, and a
per-request gate rejects any cell whose token counts differ. This
\emph{matched-work lane} measures how fast each engine executes the
same work. Its complement, the \emph{natural lane}, lets each stack
generate to its own end (Appendix~\ref{app:banks}).

\noindent\textbf{Testbed.} Each measured pair runs on one NVIDIA
A100-SXM4-80GB of a four-GPU node with no other tenant, and the two
arms alternate order across repetitions. Hardware transfer to an
H100 and an RTX A6000 follows the same protocol
(\S\ref{sec:eval-q3}, Appendix~\ref{app:h100}), with machine
details in Appendix~\ref{app:repro}.

\noindent\textbf{Metrics.} Performance experiments use SGLang's standard serving benchmark module, to replay the same seeded Poisson arrival schedule for both arms and collect metrics, with no client-side concurrency cap. We report end-to-end latency (E2E), time to first token (TTFT), and time per output token (TPOT), where TPOT is a request's mean inter-token interval. All latency means and percentiles are over requests. At the cell level, \emph{makespan} is the time from injection start until the final request completes, and \emph{drain time} is the time from the end of the injection window until completion. Whole-cell tokens per second therefore depends on both the arrival schedule and drain, rather than service capacity alone. We instead measure saturated throughput as in FaaScale~\citep{yu2026faascale}: output tokens (TPS) and completed requests (RPS) per second within the injection window at $1.25\times Q^*$ (Appendix~\ref{app:map}). Below saturation, we compare latency at equal offered load.

\noindent\textbf{Quality.} Task quality is scored with \texttt{lm-eval}~\citep{gao2024lmeval}, a unified framework for evaluating language models, on each workload's held-out documents. For GSM8K, we use a composite of its two answer
filters (Appendix~\ref{app:extraction}). To attribute quality
changes between the checkpoint and our serving, we use a
$2{\times}2$ design (Table~\ref{tab:faithfulness-2x2}, left): the base
model and the checkpoint run natively in PyTorch (arms A and B) and are served on upstream and on \sys{} with always-route (arms C and D),
all on the same documents. The paired difference-in-differences
(DiD), $(D{-}C)-(B{-}A)$, with a $95\%$ interval, isolates what
serving adds beyond the checkpoint's own cost.

\noindent\textbf{Statistics.} We repeat each main cell six times, running both arms within every repetition, and report the
mean of the per-repetition differences with a $t$-based $95\%$
interval. A difference is \emph{resolved} when its interval excludes zero, and
\emph{unresolved} otherwise, evidence of neither an effect nor
equivalence. Experiments with fewer repetitions state $n$ per table, and
single-repetition serving cells, having no interval, are never called resolved.

\subsection{Serving Comparison (Q1)}
\label{sec:eval-q1}
\label{sec:eval-e2e}

\begin{table*}[t]
  \centering
  \caption{Matched-work serving performance of \sys{} against
  upstream SGLang: mean percentage change with $95\%$ intervals
  over six paired repetitions per row. Negative is better. Makespan
  changes only through the drain after the shared injection window.}
  \label{tab:v13-headline}
  \scriptsize
  \setlength{\tabcolsep}{3pt}
\begin{tabular}{lrrrrrrr}
\toprule
& \multicolumn{2}{c}{E2E} & \multicolumn{2}{c}{TPOT} & \multicolumn{2}{c}{TTFT} & \\
\cmidrule(lr){2-3}\cmidrule(lr){4-5}\cmidrule(lr){6-7}
Rate & mean & p99 & mean & p99 & mean & p99 & makespan \\
\midrule
\multicolumn{8}{l}{\textit{GSM8K}} \\ 
$0.75\times Q^*$ & -6.6 $\pm$ 1.5 & -9.7 $\pm$ 2.8 & -6.8 $\pm$ 1.6 & -10.0 $\pm$ 3.7 & -7.1 $\pm$ 1.2 & -4.4 $\pm$ 4.8 & -0.5 $\pm$ 0.1 \\
$0.95\times Q^*$ & -36.8 $\pm$ 2.0 & -37.3 $\pm$ 2.5 & -35.6 $\pm$ 1.8 & -15.3 $\pm$ 3.8 & -66.0 $\pm$ 6.8 & -74.5 $\pm$ 3.0 & -1.4 $\pm$ 0.1 \\
$1.25\times Q^*$ & -25.7 $\pm$ 0.7 & -21.2 $\pm$ 0.6 & -10.5 $\pm$ 0.2 & -18.9 $\pm$ 1.7 & -41.5 $\pm$ 1.1 & -40.7 $\pm$ 0.9 & -7.2 $\pm$ 0.2 \\
\midrule
\multicolumn{8}{l}{\textit{BBH}} \\ 
$0.75\times Q^*$ & -4.6 $\pm$ 0.5 & -8.2 $\pm$ 1.0 & -4.4 $\pm$ 0.6 & -10.7 $\pm$ 2.7 & +4.4 $\pm$ 17.7 & +12.0 $\pm$ 37.5 & -4.7 $\pm$ 0.5 \\
$0.95\times Q^*$ & -13.6 $\pm$ 1.2 & -11.4 $\pm$ 0.3 & -14.6 $\pm$ 1.4 & -22.4 $\pm$ 1.5 & -4.9 $\pm$ 8.7 & +6.8 $\pm$ 31.3 & -7.1 $\pm$ 0.2 \\
$1.25\times Q^*$ & -8.5 $\pm$ 1.0 & -8.9 $\pm$ 0.9 & -8.8 $\pm$ 0.7 & -10.4 $\pm$ 2.0 & -5.6 $\pm$ 14.6 & -16.4 $\pm$ 7.1 & -6.3 $\pm$ 0.7 \\
\midrule
\multicolumn{8}{l}{\textit{CoQA}} \\ 
$0.75\times Q^*$ & -8.0 $\pm$ 0.6 & -10.1 $\pm$ 0.4 & +1.1 $\pm$ 5.9 & -2.0 $\pm$ 18.4 & +1.6 $\pm$ 6.8 & +5.4 $\pm$ 30.0 & -7.8 $\pm$ 0.3 \\
$0.95\times Q^*$ & -4.0 $\pm$ 4.4 & -9.2 $\pm$ 0.5 & +1.0 $\pm$ 15.1 & +1.0 $\pm$ 24.0 & -1.2 $\pm$ 5.9 & +0.8 $\pm$ 5.2 & -7.0 $\pm$ 0.4 \\
$1.25\times Q^*$ & -2.7 $\pm$ 0.2 & -9.1 $\pm$ 0.2 & -0.9 $\pm$ 0.2 & -1.7 $\pm$ 0.6 & +0.1 $\pm$ 0.6 & +0.3 $\pm$ 0.7 & -7.3 $\pm$ 0.1 \\%
 
\bottomrule
\end{tabular}
\end{table*}

We serve the released FlexiDepth checkpoint on Llama-3-8B, testing
whether \sys{} turns its native serving deficit
(\S\ref{sec:motivation}) into a gain at matched work.
Table~\ref{tab:v13-headline} reports the percentage change of each
metric against upstream, per workload and load point. Mean end-to-end latency is lower on every row, with the $95\%$ interval excluding zero on eight of nine: $\vpEtoEMeanBbhLow$ to $\vpEtoEMeanCoqaLow\%$ below the knee, $\vpEtoEMeanGsmMid\%$ at the GSM8K knee, and $\vpEtoEMeanBbhMid\%$ at the BBH knee.

The gains come from different phases, which Table~\ref{tab:ablation-body} isolates by routing one phase at a time. GSM8K's gain comes from prefill routing, which alone gives $\vpAblVpreBinarycohortEtoEMid\%$ at the knee ($n{=}1$; Appendix~\ref{app:ablation}). Shorter prefill passes also reduce concurrent decode rows' inter-token time because SGLang runs prefill and decode as separate passes on one stream. Decode routing drives BBH, matching \sys{} at the knee ($\vpAblBbhVdecFdEtoEMid\%$), with prefill contributing nothing, and produces CoQA's smaller gain. Always-route improves further at the GSM8K knee but becomes slower than upstream at the CoQA knee ($\vpAblCoqaIntegratedAlwaysskipEtoEMid\%$), while disabling promotion reduces the GSM8K knee gain to $\vpAblVskipperNoupgradeEtoEMid\%$.

\IfFileExists{generated/ablation_rows.tex}{%
\begin{table}[t]
  \centering
   \caption{GSM8K mechanism ablation: mean E2E and makespan change (\%) against upstream. The served arm uses $n{=}6$ repetitions and each ablation arm $n{=}1$. All workloads appear in Appendix~\ref{app:ablation}.}
  \label{tab:ablation-body}
  \scriptsize
  \setlength{\tabcolsep}{3pt}
  \begin{tabular}{lrrrrrrrrrr}
  \toprule
  & \multicolumn{2}{c}{served ($n{=}6$)} & \multicolumn{2}{c}{decode only} & \multicolumn{2}{c}{prefill only} & \multicolumn{2}{c}{always route} & \multicolumn{2}{c}{no promotion} \\
  rate & E2E & makespan & E2E & makespan & E2E & makespan & E2E & makespan & E2E & makespan \\
  \midrule
  0.75$\times$ & -6.6 $\pm$ 1.5 & -0.5 $\pm$ 0.1 & +0.4 & +0.0 & -8.6 & -0.7 & -5.5 & -1.0 & -6.7 & -0.5 \\
0.95$\times$ & -36.8 $\pm$ 2.0 & -1.4 $\pm$ 0.1 & -14.3 & -0.9 & -35.6 & -1.2 & -46.5 & -1.9 & -32.6 & -1.2 \\
1.25$\times$ & -25.7 $\pm$ 0.7 & -7.2 $\pm$ 0.2 & -7.8 & -2.6 & -19.9 & -5.4 & -26.3 & -7.4 & -24.5 & -7.0 \\
 
  \bottomrule
  \end{tabular}
\end{table}}{}

Near saturation, queueing amplifies execution savings into large
latency gains (Appendix~\ref{app:map}). At the same offered load, requests
wait less: time to first token falls by $\vpTTFTMeanGsmMidAbs\%$ at the
GSM8K knee, where shorter prefill passes clear the admission queue.
Under saturation at $1.25\times Q^*$, \sys{} raises RPS by
$\vpCapGsmRpsAbs\%$ on GSM8K and $\vpCapBbhRpsAbs\%$ on BBH. CoQA, bound by prefill that \sys{} tends to avoid routing, shows no resolved change (Table~\ref{tab:capacity}).

\begin{table}[t]
  \centering
  \caption{Throughput under saturation at $1.25\times Q^*$: mean change (\%) with
  $95\%$ intervals against upstream in TPS and RPS inside the injection
  window; positive is better. GSM8K, BBH, CoQA: Llama-3-8B on the A100
  ($n{=}6$); other columns: GSM8K ($n{=}3$).}
  \label{tab:capacity}
  \scriptsize
  \setlength{\tabcolsep}{3pt}
  \begin{tabular}{lrrrrrrr}
  \toprule
  & GSM8K & BBH & CoQA & Qwen3-4B & Qwen3-8B & H100 & RTX A6000 \\
  \midrule
  \IfFileExists{generated/capacity_rows.tex}{TPS & $+10.4 \pm 0.3$ & $+4.1 \pm 0.4$ & $-1.0 \pm 1.6$ & $-0.2 \pm 0.3$ & $+2.5 \pm 0.2$ & $+18.5 \pm 3.0$ & $+9.1 \pm 2.1$ \\
RPS & $+11.3 \pm 0.2$ & $+7.4 \pm 0.3$ & $-1.0 \pm 1.7$ & $-0.2 \pm 0.4$ & $+2.8 \pm 0.1$ & $+19.3 \pm 3.4$ & $+9.3 \pm 1.7$ \\
 }{}
  \bottomrule
  \end{tabular}
\end{table}

CoQA is tail-heavy: $\vpNatLenUpRunawayPctCoqaLow\%$ of the base
model's generations run past $4{,}096$ tokens and carry
$\vpNatLenUpRunawayTokPctCoqaLow\%$ of a cell's output tokens.
\sys{} finishes this tail faster, improving makespan by
$\vpMakespanCoqaMidAbs$--$\vpMakespanCoqaLowAbs\%$ at every rate; the
gain sits in the long requests (Appendix~\ref{app:tails}).

At matched work, \sys{} thus converts FlexiDepth's serving deficit
into lower mean and p99 end-to-end latency across the grid, largest at and
beyond the knee on GSM8K and BBH.

\subsection{Task Quality (Q2)}
\label{sec:eval-q2}

\begin{table}[t]
  \centering
  \caption{Task quality (\%; GSM8K composite exact match, BBH exact match,
  CoQA F1). Left: base (A) and checkpoint (B) in PyTorch, served on upstream
  (C) and always-route \sys{} (D), with the DiD. Right: served at the knee; routed: share of decode
  rows, hybrid / always-route.}
  \label{tab:faithfulness-2x2}
  \scriptsize
  \setlength{\tabcolsep}{3pt}
\begin{tabular}{lrrrrrrrrrrr}
\toprule
& \multicolumn{7}{c}{PyTorch versus served (\texttt{lm-eval})} & \multicolumn{4}{c}{Served at the knee} \\
\cmidrule(lr){2-8}\cmidrule(lr){9-12}
Task & A & B & C & D & $(B{-}A)$ & $(D{-}C)$ & DiD (pp) & upstream & hybrid & always-route & routed (\%) \\
\midrule
\IfFileExists{generated/quality_merged_rows.tex}{GSM8K & 77.41 & 74.15 & 77.79 & 74.30 & -3.26 & -3.49 & -0.23 $\pm$ 0.68 & 77.63 & 74.00 & 74.07 & 90 / 100 \\
CoQA & 78.31 & 78.99 & 78.31 & 79.07 & +0.68 & +0.75 & +0.07 $\pm$ 0.12 & 76.74 & 77.68 & 77.73 & 71 / 100 \\
BBH & 68.04 & 59.65 & 67.93 & 59.99 & -8.39 & -7.94 & +0.45 $\pm$ 0.55 & 67.87 & 59.47 & 59.68 & 99 / 100 \\
 }{}
\bottomrule
\end{tabular}
\end{table}

Serving adds no resolved quality loss beyond the checkpoint's own.
Table~\ref{tab:faithfulness-2x2} (left) reports the $2{\times}2$: the
DiD is $\vpQGsmDod \pm \vpQGsmDodCi$~pp on GSM8K, $\vpQBbhDod \pm
\vpQBbhDodCi$~pp on BBH and $\vpQCoqaDod \pm \vpQCoqaDodCi$~pp on
CoQA, each interval containing zero. The checkpoint's own change,
$B{-}A$, is the skipper's quality--compute trade-off. The same serving interface can therefore support checkpoints with different trade-offs.

Table~\ref{tab:faithfulness-2x2} (right) evaluates the served hybrid under load at the knee. It shows no resolved difference from always-route: $\vpLqGsmMidDelta \pm \vpLqGsmMidDeltaCi$~pp on GSM8K, $\vpLqBbhMidDelta \pm \vpLqBbhMidDeltaCi$~pp on BBH, and $\vpLqCoqaMidDelta \pm \vpLqCoqaMidDeltaCi$~pp on CoQA. Intervals are computed per document with $n{=}1$, and other load rates appear in Appendix~\ref{app:loaded-sweep}.

\subsection{Generality (Q3)}
\label{sec:eval-q3}
The framework generalizes across skippers, model families, workloads, and hardware. New skippers, models, and devices
reuse the same interface and execution framework without scheduler or
memory-manager redesign or per-workload tuning, aside from architecture-specific projections.

\IfFileExists{figures/sweep_heatmap_v2_mean.pdf}{%
\begin{figure}[t]
  \centering
  \begin{minipage}[t]{0.30\linewidth}\centering\includegraphics[width=\linewidth]{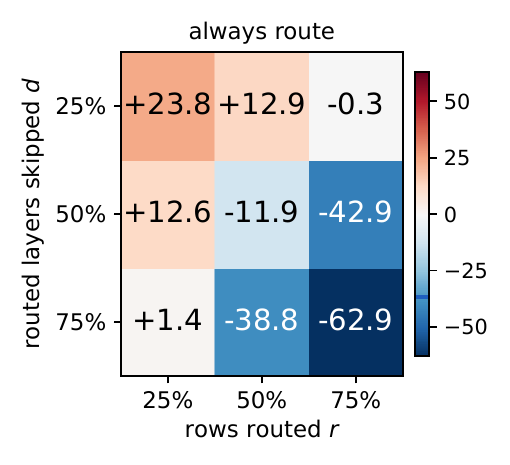}\\[-2pt]{\scriptsize (a) always route}\end{minipage}%
  \hfill%
  \begin{minipage}[t]{0.30\linewidth}\centering\includegraphics[width=\linewidth]{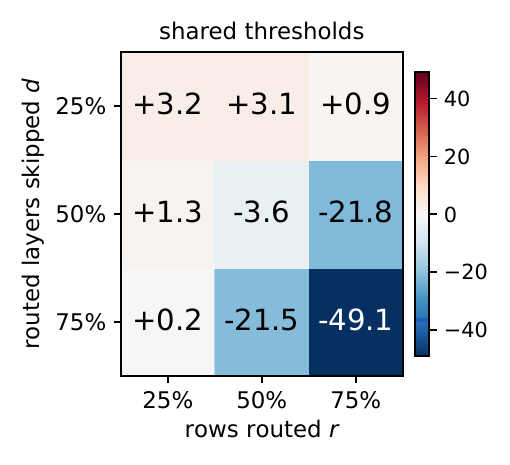}\\[-2pt]{\scriptsize (b) shared thresholds}\end{minipage}%
  \hfill%
  \begin{minipage}[t]{0.30\linewidth}\centering\includegraphics[width=\linewidth]{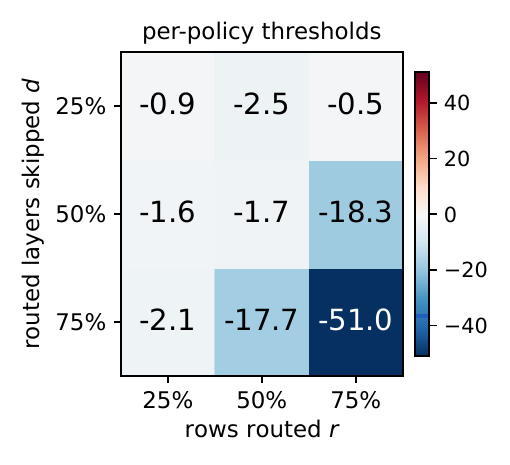}\\[-2pt]{\scriptsize (c) per-policy thresholds}\end{minipage}%
  \caption{Mean E2E change (\%) against upstream for the nine
  RandomSkip arms at the GSM8K knee ($n{=}1$), by rows routed $r$
  and routed layers skipped $d$, under the three switch settings
  (a--c). 
  }
  \label{fig:sweep}
\end{figure}}{}

\noindent\textbf{Skippers.} Alongside FlexiDepth, we define a deterministic
mock skipper (RandomSkip) that implements the interface by routing a fixed
fraction $r$ of rows around a fixed fraction $d$ of routed layers,
giving skip ratio $s=r\,d$. Figure~\ref{fig:sweep}
maps the nine arms' mean E2E change under three switch settings: always route (a), FlexiDepth's thresholds for every arm (b), and rule-derived thresholds per policy (c). Without the switch, 
unprofitable arms cost up to $\vpSweepUngEtoEMeanWorst\%$. 
With shared thresholds, routing
becomes profitable between $s = \tfrac{3}{16}$ and $s = \tfrac14$,
where the rule places the crossover (Appendix~\ref{app:roofline}),
and gains reach $\vpSweepEtoEMeanBest\%$ at $r{=}d{=}75\%$. Per-policy
thresholds keep unprofitable arms dense while retaining gains for profitable ones.

\begin{table}[t]
  \centering
  \caption{Cross-model comparison on GSM8K at each model's knee.
Matched-work mean E2E uses $n{=}6$ for Llama and $n{=}3$ for Qwen. Served accuracy uses one repetition. Skip is the always-route
Project-Only decode share.}
  \label{tab:models}
  \scriptsize
  \setlength{\tabcolsep}{4pt}
  \begin{tabular}{lrrrrrrr}
  \toprule
  & & \multicolumn{3}{c}{mean E2E (s)} & \multicolumn{3}{c}{accuracy (\%)} \\
  \cmidrule(lr){3-5}\cmidrule(lr){6-8}
  Model & skip & upstream & \sys{} & change (\%) & upstream & \sys{} & change (pp) \\
  \midrule
  Llama-3-8B (FlexiDepth) & $\vpLlamaKneeSDecode$ & $\vpPMLlamaGsmLatUp$ & $\vpPMLlamaGsmLatHyb$ & $\vpPMLlamaGsmLat \pm \vpPMLlamaGsmLatCi$ & $\vpPMLlamaGsmQUp$ & $\vpPMLlamaGsmQHyb$ & $\vpPMLlamaGsmQ \pm \vpPMLlamaGsmQCi$ \\
  Qwen3-4B & $\vpQwenFourKneeSDecode$ & $\vpPMQwenFourLatUp$ & $\vpPMQwenFourLatHyb$ & $\vpPMQwenFourLat \pm \vpPMQwenFourLatCi$ & $\vpPMQwenFourQUp$ & $\vpPMQwenFourQHyb$ & $\vpPMQwenFourQ \pm \vpPMQwenFourQCi$ \\
  Qwen3-8B & $\vpQwenEightKneeSDecode$ & $\vpPMQwenEightLatUp$ & $\vpPMQwenEightLatHyb$ & $\vpPMQwenEightLat \pm \vpPMQwenEightLatCi$ & $\vpPMQwenEightQUp$ & $\vpPMQwenEightQHyb$ & $\vpPMQwenEightQ \pm \vpPMQwenEightQCi$ \\
  \bottomrule
  \end{tabular}
\end{table}

\noindent\textbf{Models.} We train Qwen3 skippers with only the first of FlexiDepth's two training
stages (Appendix~\ref{app:qwen}). They use the same RUN/Project-Only interface
and execution path as FlexiDepth on Llama-3-8B. The only model-specific
change is a hook that applies Qwen3's per-head query and key normalization
when a skipped layer writes its KV, as Qwen3's own attention does; the
switching thresholds follow from the rule in \S\ref{sec:design-engagement}.
Table~\ref{tab:models} pairs latency and quality at each model's
knee. Qwen3-4B, whose always-route decode skip of $\vpQwenFourKneeSDecode$ is
too low for routing to be profitable at these loads, shows no resolved change in
mean E2E at any load ($\vpQwenFourEtoEMeanQwenFourMid \pm
\vpQwenFourEtoEMeanQwenFourMidCi\%$ at the knee).
Qwen3-8B, at skip $\vpQwenEightKneeSDecode$, lowers mean
E2E by $\vpQwenEightEtoEMeanQwenEightMidAbs \pm
\vpQwenEightEtoEMeanQwenEightMidCi\%$ with no resolved quality loss beyond
the checkpoint's own (Appendix~\ref{app:qwen}). Across three Qwen3-8B checkpoints,
serving gains grow with the skip rate, and along the lower-penalty run, continued training raised skipping and accuracy together, suggesting
skipper training can further improve this quality--efficiency
trade-off.

\begin{table*}[!tb]
  \centering
  \caption{Hardware transfer: the GSM8K matched-work comparison on
  an H100 ($Q^* = \vpLadderHundredGsmKnee$~req/s) and an RTX A6000
  ($Q^* = \vpLadderAsixGsmKnee$~req/s), as mean percentage change
  with $95\%$ intervals over three paired repetitions against
  upstream on the same device.}
  \label{tab:transfer}
  \label{tab:h100}
  \label{tab:a6000}
  \scriptsize
  \setlength{\tabcolsep}{3pt}
\begin{tabular}{llrrrrrrrr}
\toprule
& & & \multicolumn{2}{c}{E2E} & \multicolumn{2}{c}{TPOT} & \multicolumn{2}{c}{TTFT} & \\
\cmidrule(lr){4-5}\cmidrule(lr){6-7}\cmidrule(lr){8-9}
Device & Rate & $n$ & mean & p99 & mean & p99 & mean & p99 & makespan \\
\midrule
\IfFileExists{generated/h100_rows.tex}{H100 & $0.75\times Q^*$ & 3 & -6.0 $\pm$ 3.2 & -8.8 $\pm$ 5.2 & -6.3 $\pm$ 3.5 & -9.8 $\pm$ 9.1 & -5.2 $\pm$ 0.2 & -9.8 $\pm$ 6.4 & -0.5 $\pm$ 0.3 \\
H100 & $0.95\times Q^*$ & 3 & -39.4 $\pm$ 0.2 & -34.2 $\pm$ 9.6 & -31.4 $\pm$ 7.6 & -11.5 $\pm$ 7.9 & -91.7 $\pm$ 4.8 & -93.0 $\pm$ 2.2 & -2.3 $\pm$ 2.2 \\
H100 & $1.25\times Q^*$ & 3 & -35.8 $\pm$ 3.4 & -32.1 $\pm$ 3.0 & -15.7 $\pm$ 1.5 & -24.4 $\pm$ 9.3 & -51.8 $\pm$ 3.8 & -50.8 $\pm$ 2.4 & -9.8 $\pm$ 1.1 \\%
 }{}
\IfFileExists{generated/a6000_rows.tex}{RTX A6000 & $0.75\times Q^*$ & 3 & -1.4 $\pm$ 2.3 & -7.3 $\pm$ 2.4 & -1.4 $\pm$ 2.5 & -9.3 $\pm$ 2.5 & -8.9 $\pm$ 2.3 & -16.4 $\pm$ 8.9 & +0.3 $\pm$ 0.7 \\
RTX A6000 & $0.95\times Q^*$ & 3 & -30.0 $\pm$ 4.2 & -27.7 $\pm$ 4.6 & -23.0 $\pm$ 5.0 & -12.0 $\pm$ 6.0 & -81.1 $\pm$ 2.8 & -62.9 $\pm$ 4.7 & -2.2 $\pm$ 0.2 \\
RTX A6000 & $1.25\times Q^*$ & 3 & -27.1 $\pm$ 4.2 & -26.6 $\pm$ 4.4 & -9.9 $\pm$ 1.3 & -15.8 $\pm$ 5.5 & -34.8 $\pm$ 5.5 & -33.3 $\pm$ 5.4 & -6.7 $\pm$ 1.1 \\%
 }{}
\bottomrule
\end{tabular}
\end{table*}

\noindent\textbf{Hardware.} We run the same implementation on an H100
and an RTX A6000 without workload-specific tuning. Each device uses
rule-derived thresholds and its own knee, while only device-specific
kernel tiles and, on the A6000, the decode-graph range change
(Table~\ref{tab:transfer}). At the knees, mean E2E falls by
$\vpHoneEtoEMeanGsmMidAbs \pm \vpHoneEtoEMeanGsmMidCi\%$ on the H100
and $\vpAsixEtoEMeanGsmMidAbs \pm \vpAsixEtoEMeanGsmMidCi\%$ on the
A6000 (GSM8K, $n{=}3$; Appendix~\ref{app:h100}).

\section{Related Work}
\label{sec:related}

\noindent\textbf{Layer skipping.} Prior work reduces transformer depth by selecting layers per token~\citep{luo2025flexidepth,yang2025dash}, skipping sublayers~\citep{he2025adaskip}, using fixed or temporally scheduled layer subsets~\citep{liu2024unified, rajabzadeh2026loradrop}, exiting early~\citep{schuster2022calm,elhoushi2024layerskip,chen2024eellm}, or removing layers~\citep{men2024shortgpt}. Other methods constrain the policy to simplify execution~\citep{corro2023skipdecode,raposo2024mixtureofdepths}. These works primarily decide \emph{what} to skip, and released FlexiDepth, AdaSkip, and LoRA-Drop implementations use specialized PyTorch paths. \sys{} instead enables layer skipping in modern serving engines by providing a RUN/Project-Only interface.

\noindent\textbf{Serving dynamic computation.} Modern engines~\citep{yu2022orca,kwon2023pagedattention,zheng2024sglang,agrawal2024sarathi} schedule at token granularity but execute each batch through the full model. DREX~\citep{liu2025drex} and Apparate~\citep{dai2024apparate} serve early exits, where each decision removes a suffix of layers; DREX rebatches by depth and transfers state across depths. Multi-adapter LoRA serving~\citep{hu2022lora,chen2024punica,sheng2024slora} separate customized model adapter from batched execution, but retain fixed layer traversal. MoE systems~\citep{moe} route tokens across experts within a layer, whereas \sys{} routes across depth while keeping the batch and captured graph fixed and preserving per-layer KV state.

\section{Conclusion and Future Work}
\label{sec:limitations}
\label{sec:conclusion}

We present \sys{}, a serving virtualization layer that translates per-token interior layer skipping into gains in a modern LLM serving engine. Its RUN/Project-Only interface, mixed-depth cohort execution, and profitability-aware switching preserve the batching, cache, and graph optimizations of SGLang. \sys{} converts FlexiDepth's saved computation into lower latency and higher saturated throughput without statistically resolved serving-induced quality loss, and the framework extends across additional policies, Qwen3 skippers, workloads, and three GPUs without workload-specific tuning. 
Three directions could further broaden \sys{}'s scope. (1) Extend execution support across additional serving engines and attention backends. (2) Generalize the interface to sublayer skippers such as AdaSkip and more model families. (3) Support distributed serving. Separate prefill and decode modes naturally fit disaggregation~\citep{zhong2024distserve}, while tensor and pipeline parallelism require routing-aware execution across devices. Pipeline parallelism is particularly promising: assigning stages using profiled skip rates could balance dynamic work and translate layer skipping into pipeline-level gains.

\clearpage
\section*{Reproducibility statement}
The main code repository is available anonymously at
\textcolor{blue}{\url{https://github.com/AKafakA/sglang-vskipper/tree/vskipper-ref}}
The \sys{} implementation over SGLang, the training scripts for the
Qwen3 skippers, and the experiment scripts that reproduce the
results, from serving cells to measurement-derived tables, figures
and numerical summaries. The two Qwen3 skippers we trained are
released at \textcolor{blue}{\url{https://huggingface.co/asdwb/vskip-flexidepth-qwen-3}}.

GPU reproduction uses the repository scripts with the model,
dataset and checkpoint assets specified in Appendix~\ref{app:repro}
and the code guides. The serving cells take about $75$ GPU-hours
across the three devices, under USD~$100$ at the recorded prices.
Each Qwen3 skipper trains in $40$ to $60$ A100-hours.
Appendix~\ref{app:repro} gives the revisions, launch profile, $Q^*$ sweeps, training setup and cost breakdown.
Appendix~\ref{app:gates} lists the checks each cell passed;
Appendix~\ref{app:plugin} gives a policy-plugin example.


\bibliographystyle{ACM-Reference-Format}
\bibliography{references}

@inproceedings{luo2025flexidepth,
title={Adaptive Layer-skipping in Pre-trained {LLM}s},
author={Xuan Luo and Weizhi Wang and Xifeng Yan},
booktitle={Second Conference on Language Modeling},
year={2025},
url={https://openreview.net/forum?id=Gu0XSax2YS}
}

@misc{rajabzadeh2026loradrop,
      title={LoRA-Drop: Temporal LoRA Decoding for Efficient LLM Inference}, 
      author={Hossein Rajabzadeh and Maryam Dialameh and Chul B. Park and Il-Min Kim and Hyock Ju Kwon},
      year={2026},
      eprint={2601.02569},
      archivePrefix={arXiv},
      primaryClass={cs.CL},
      url={https://arxiv.org/abs/2601.02569}, 
}

@misc{liu2025drex,
      title={Dynamic Rebatching for Efficient Early-Exit Inference with DREX}, 
      author={Xuting Liu and Daniel Alexander and Siva Kesava Reddy Kakarla and Behnaz Arzani and Vincent Liu},
      year={2025},
      eprint={2512.15705},
      archivePrefix={arXiv},
      primaryClass={cs.DC},
      url={https://arxiv.org/abs/2512.15705}, 
}

@inproceedings{elhoushi2024layerskip,
    title = "{L}ayer{S}kip: Enabling Early Exit Inference and Self-Speculative Decoding",
    author = "Elhoushi, Mostafa  and
      Shrivastava, Akshat  and
      Liskovich, Diana  and
      Hosmer, Basil  and
      Wasti, Bram  and
      Lai, Liangzhen  and
      Mahmoud, Anas  and
      Acun, Bilge  and
      Agarwal, Saurabh  and
      Roman, Ahmed  and
      Aly, Ahmed  and
      Chen, Beidi  and
      Wu, Carole-Jean",
    editor = "Ku, Lun-Wei  and
      Martins, Andre  and
      Srikumar, Vivek",
    booktitle = "Proceedings of the 62nd Annual Meeting of the Association for Computational Linguistics (Volume 1: Long Papers)",
    month = aug,
    year = "2024",
    address = "Bangkok, Thailand",
    publisher = "Association for Computational Linguistics",
    url = "https://aclanthology.org/2024.acl-long.681/",
    doi = "10.18653/v1/2024.acl-long.681",
    pages = "12622--12642"
}

@inproceedings{zheng2024sglang,
author = {Zheng, Lianmin and Yin, Liangsheng and Xie, Zhiqiang and Sun, Chuyue and Huang, Jeff and Yu, Cody Hao and Cao, Shiyi and Kozyrakis, Christos and Stoica, Ion and Gonzalez, Joseph E. and Barrett, Clark and Sheng, Ying},
title = {SGLang: efficient execution of structured language model programs},
year = {2024},
isbn = {9798331314385},
publisher = {Curran Associates Inc.},
address = {Red Hook, NY, USA},
booktitle = {Proceedings of the 38th International Conference on Neural Information Processing Systems},
articleno = {2000},
numpages = {27},
location = {Vancouver, BC, Canada},
series = {NIPS '24}
}

@inproceedings{kwon2023pagedattention,
author = {Kwon, Woosuk and Li, Zhuohan and Zhuang, Siyuan and Sheng, Ying and Zheng, Lianmin and Yu, Cody Hao and Gonzalez, Joseph and Zhang, Hao and Stoica, Ion},
title = {Efficient Memory Management for Large Language Model Serving with PagedAttention},
year = {2023},
isbn = {9798400702297},
publisher = {Association for Computing Machinery},
address = {New York, NY, USA},
url = {https://doi.org/10.1145/3600006.3613165},
doi = {10.1145/3600006.3613165},
booktitle = {Proceedings of the 29th Symposium on Operating Systems Principles},
pages = {611–626},
numpages = {16},
location = {Koblenz, Germany},
series = {SOSP '23}
}

@inproceedings {yu2022orca,
author = {Gyeong-In Yu and Joo Seong Jeong and Geon-Woo Kim and Soojeong Kim and Byung-Gon Chun},
title = {Orca: A Distributed Serving System for {Transformer-Based} Generative Models},
booktitle = {16th USENIX Symposium on Operating Systems Design and Implementation (OSDI 22)},
year = {2022},
isbn = {978-1-939133-28-1},
address = {Carlsbad, CA},
pages = {521--538},
url = {https://www.usenix.org/conference/osdi22/presentation/yu},
publisher = {USENIX Association},
month = jul
}

@misc{corro2023skipdecode,
      title={SkipDecode: Autoregressive Skip Decoding with Batching and Caching for Efficient LLM Inference}, 
      author={Luciano Del Corro and Allie Del Giorno and Sahaj Agarwal and Bin Yu and Ahmed Awadallah and Subhabrata Mukherjee},
      year={2023},
      eprint={2307.02628},
      archivePrefix={arXiv},
      primaryClass={cs.CL},
      url={https://arxiv.org/abs/2307.02628}, 
}

@misc{raposo2024mixtureofdepths,
      title={Mixture-of-Depths: Dynamically allocating compute in transformer-based language models}, 
      author={David Raposo and Sam Ritter and Blake Richards and Timothy Lillicrap and Peter Conway Humphreys and Adam Santoro},
      year={2024},
      eprint={2404.02258},
      archivePrefix={arXiv},
      primaryClass={cs.LG},
      url={https://arxiv.org/abs/2404.02258}, 
}

@inproceedings{schuster2022calm,
author = {Schuster, Tal and Fisch, Adam and Gupta, Jai and Dehghani, Mostafa and Bahri, Dara and Tran, Vinh Q. and Tay, Yi and Metzler, Donald},
title = {Confident adaptive language modeling},
year = {2022},
isbn = {9781713871088},
publisher = {Curran Associates Inc.},
address = {Red Hook, NY, USA},
booktitle = {Proceedings of the 36th International Conference on Neural Information Processing Systems},
articleno = {1269},
numpages = {17},
location = {New Orleans, LA, USA},
series = {NIPS '22}
}

@inproceedings{dai2024apparate,
author = {Dai, Yinwei and Pan, Rui and Iyer, Anand and Li, Kai and Netravali, Ravi},
title = {Apparate: Rethinking Early Exits to Tame Latency-Throughput Tensions in ML Serving},
year = {2024},
isbn = {9798400712517},
publisher = {Association for Computing Machinery},
address = {New York, NY, USA},
url = {https://doi.org/10.1145/3694715.3695963},
doi = {10.1145/3694715.3695963},
booktitle = {Proceedings of the ACM SIGOPS 30th Symposium on Operating Systems Principles},
pages = {607–623},
numpages = {17},
location = {Austin, TX, USA},
series = {SOSP '24}
}

@inproceedings{agrawal2024sarathi,
author = {Agrawal, Amey and Kedia, Nitin and Panwar, Ashish and Mohan, Jayashree and Kwatra, Nipun and Gulavani, Bhargav S. and Tumanov, Alexey and Ramjee, Ramachandran},
title = {Taming throughput-latency tradeoff in LLM inference with sarathi-serve},
year = {2024},
isbn = {978-1-939133-40-3},
publisher = {USENIX Association},
address = {USA},
booktitle = {Proceedings of the 18th USENIX Conference on Operating Systems Design and Implementation},
articleno = {7},
numpages = {18},
location = {Santa Clara, CA, USA},
series = {OSDI'24}
}

@inproceedings{chen2024eellm,
author = {Chen, Yanxi and Pan, Xuchen and Li, Yaliang and Ding, Bolin and Zhou, Jingren},
title = {EE-LLM: large-scale training and inference of early-exit large language models with 3D parallelism},
year = {2024},
publisher = {JMLR.org},
booktitle = {Proceedings of the 41st International Conference on Machine Learning},
articleno = {277},
numpages = {27},
location = {Vienna, Austria},
series = {ICML'24}
}

@inproceedings{men2024shortgpt,
    title = "{S}hort{GPT}: Layers in Large Language Models are More Redundant Than You Expect",
    author = "Men, Xin  and
      Xu, Mingyu  and
      Zhang, Qingyu  and
      Yuan, Qianhao  and
      Wang, Bingning  and
      Lin, Hongyu  and
      Lu, Yaojie  and
      Han, Xianpei  and
      Chen, Weipeng",
    editor = "Che, Wanxiang  and
      Nabende, Joyce  and
      Shutova, Ekaterina  and
      Pilehvar, Mohammad Taher",
    booktitle = "Findings of the Association for Computational Linguistics: ACL 2025",
    month = jul,
    year = "2025",
    address = "Vienna, Austria",
    publisher = "Association for Computational Linguistics",
    url = "https://aclanthology.org/2025.findings-acl.1035/",
    doi = "10.18653/v1/2025.findings-acl.1035",
    pages = "20192--20204",
    ISBN = "979-8-89176-256-5"
}

@inproceedings{he2025adaskip,
author = {He, Zhuomin and Yao, Yizhen and Zuo, Pengfei and Gao, Bin and Li, Qinya and Zheng, Zhenzhe and Wu, Fan},
title = {AdaSkip: adaptive sublayer skipping for accelerating long-context LLM inference},
year = {2025},
isbn = {978-1-57735-897-8},
publisher = {AAAI Press},
url = {https://doi.org/10.1609/aaai.v39i22.34579},
doi = {10.1609/aaai.v39i22.34579},
booktitle = {Proceedings of the Thirty-Ninth AAAI Conference on Artificial Intelligence and Thirty-Seventh Conference on Innovative Applications of Artificial Intelligence and Fifteenth Symposium on Educational Advances in Artificial Intelligence},
articleno = {2681},
numpages = {9},
series = {AAAI'25/IAAI'25/EAAI'25}
}

@misc{yang2025dash,
      title={DASH: Input-Aware Dynamic Layer Skipping for Efficient LLM Inference with Markov Decision Policies}, 
      author={Ning Yang and Fangxin Liu and Junjie Wang and Tao Yang and Kan Liu and Haibing Guan and Li Jiang},
      year={2025},
      eprint={2505.17420},
      archivePrefix={arXiv},
      primaryClass={cs.CL},
      url={https://arxiv.org/abs/2505.17420}, 
}

@misc{liu2024unified,
      title={Accelerating Inference in Large Language Models with a Unified Layer Skipping Strategy}, 
      author={Yijin Liu and Fandong Meng and Jie Zhou},
      year={2024},
      eprint={2404.06954},
      archivePrefix={arXiv},
      primaryClass={cs.CL},
      url={https://arxiv.org/abs/2404.06954}, 
}

@misc{agrawal2023sarathi,
      title={SARATHI: Efficient LLM Inference by Piggybacking Decodes with Chunked Prefills}, 
      author={Amey Agrawal and Ashish Panwar and Jayashree Mohan and Nipun Kwatra and Bhargav S. Gulavani and Ramachandran Ramjee},
      year={2023},
      eprint={2308.16369},
      archivePrefix={arXiv},
      primaryClass={cs.LG},
      url={https://arxiv.org/abs/2308.16369}, 
}

@inproceedings{
hu2022lora,
title={Lo{RA}: Low-Rank Adaptation of Large Language Models},
author={Edward J Hu and yelong shen and Phillip Wallis and Zeyuan Allen-Zhu and Yuanzhi Li and Shean Wang and Lu Wang and Weizhu Chen},
booktitle={International Conference on Learning Representations},
year={2022},
url={https://openreview.net/forum?id=nZeVKeeFYf9}
}

@misc{sheng2024slora,
      title={S-LoRA: Serving Thousands of Concurrent LoRA Adapters}, 
      author={Ying Sheng and Shiyi Cao and Dacheng Li and Coleman Hooper and Nicholas Lee and Shuo Yang and Christopher Chou and Banghua Zhu and Lianmin Zheng and Kurt Keutzer and Joseph E. Gonzalez and Ion Stoica},
      year={2024},
      eprint={2311.03285},
      archivePrefix={arXiv},
      primaryClass={cs.LG},
      url={https://arxiv.org/abs/2311.03285}, 
}

@misc{chen2024punica,
      title={Punica: Multi-Tenant LoRA Serving}, 
      author={Lequn Chen and Zihao Ye and Yongji Wu and Danyang Zhuo and Luis Ceze and Arvind Krishnamurthy},
      year={2023},
      eprint={2310.18547},
      archivePrefix={arXiv},
      primaryClass={cs.DC},
      url={https://arxiv.org/abs/2310.18547}, 
}

@article{williams2009roofline,
author = {Williams, Samuel and Waterman, Andrew and Patterson, David},
title = {Roofline: an insightful visual performance model for multicore architectures},
year = {2009},
issue_date = {April 2009},
publisher = {Association for Computing Machinery},
address = {New York, NY, USA},
volume = {52},
number = {4},
issn = {0001-0782},
url = {https://doi.org/10.1145/1498765.1498785},
doi = {10.1145/1498765.1498785},
journal = {Commun. ACM},
month = apr,
pages = {65–76},
numpages = {12}
}

@misc{cobbe2021gsm8k,
      title={Training Verifiers to Solve Math Word Problems}, 
      author={Karl Cobbe and Vineet Kosaraju and Mohammad Bavarian and Mark Chen and Heewoo Jun and Lukasz Kaiser and Matthias Plappert and Jerry Tworek and Jacob Hilton and Reiichiro Nakano and Christopher Hesse and John Schulman},
      year={2021},
      eprint={2110.14168},
      archivePrefix={arXiv},
      primaryClass={cs.LG},
      url={https://arxiv.org/abs/2110.14168}, 
}

@misc{suzgun2023bbh,
      title={Challenging BIG-Bench Tasks and Whether Chain-of-Thought Can Solve Them}, 
      author={Mirac Suzgun and Nathan Scales and Nathanael Schärli and Sebastian Gehrmann and Yi Tay and Hyung Won Chung and Aakanksha Chowdhery and Quoc V. Le and Ed H. Chi and Denny Zhou and Jason Wei},
      year={2022},
      eprint={2210.09261},
      archivePrefix={arXiv},
      primaryClass={cs.CL},
      url={https://arxiv.org/abs/2210.09261}, 
}

@article{reddy2019coqa,
    title = "{C}o{QA}: A Conversational Question Answering Challenge",
    author = "Reddy, Siva  and
      Chen, Danqi  and
      Manning, Christopher D.",
    editor = "Lee, Lillian  and
      Johnson, Mark  and
      Roark, Brian  and
      Nenkova, Ani",
    journal = "Transactions of the Association for Computational Linguistics",
    volume = "7",
    year = "2019",
    address = "Cambridge, MA",
    publisher = "MIT Press",
    url = "https://aclanthology.org/Q19-1016/",
    doi = "10.1162/tacl_a_00266",
    pages = "249--266"
}

@misc{gao2024lmeval,
  author       = {Gao, Leo and Tow, Jonathan and Abbasi, Baber and Biderman, Stella and Black, Sid and DiPofi, Anthony and Foster, Charles and Golding, Laurence and Hsu, Jeffrey and Le Noac'h, Alain and Li, Haonan and McDonell, Kyle and Muennighoff, Niklas and Ociepa, Chris and Phang, Jason and Reynolds, Laria and Schoelkopf, Hailey and Skowron, Aviya and Sutawika, Lintang and Tang, Eric and Thite, Anish and Wang, Ben and Wang, Kevin and Zou, Andy},
  title        = {The Language Model Evaluation Harness},
  month        = 07,
  year         = 2024,
  publisher    = {Zenodo},
  version      = {v0.4.3},
  doi          = {10.5281/zenodo.12608602},
  url          = {https://zenodo.org/records/12608602}
}

@inproceedings{
yu2026faascale,
title={FaaScale: Unlocking Fast {LLM} Scaling for Serverless Inference},
author={Minchen Yu and Rui Yang and Chaobo Jia and Zhaoyuan Su and Sheng Yao and Tingfeng Lan and Yuchen Yang and Zirui Wang and Yue Cheng and Wei Wang and Ao Wang and Ruichuan Chen},
booktitle={Ninth Conference on Machine Learning and Systems},
year={2026},
url={https://openreview.net/forum?id=jgL8LuOVyT}
}

@inproceedings{zhong2024distserve,
author = {Zhong, Yinmin and Liu, Shengyu and Chen, Junda and Hu, Jianbo and Zhu, Yibo and Liu, Xuanzhe and Jin, Xin and Zhang, Hao},
title = {DistServe: disaggregating prefill and decoding for goodput-optimized large language model serving},
year = {2024},
isbn = {978-1-939133-40-3},
publisher = {USENIX Association},
address = {USA},
booktitle = {Proceedings of the 18th USENIX Conference on Operating Systems Design and Implementation},
articleno = {11},
numpages = {18},
location = {Santa Clara, CA, USA},
series = {OSDI'24}
}

@misc{moe,
      title={Outrageously Large Neural Networks: The Sparsely-Gated Mixture-of-Experts Layer}, 
      author={Noam Shazeer and Azalia Mirhoseini and Krzysztof Maziarz and Andy Davis and Quoc Le and Geoffrey Hinton and Jeff Dean},
      year={2017},
      eprint={1701.06538},
      archivePrefix={arXiv},
      primaryClass={cs.LG},
      url={https://arxiv.org/abs/1701.06538}, 
}

\appendix

\section{A Minimal Skipper Plugin}
\label{app:plugin}

A policy plugin returns one action per row without executing either
branch. The projector specifies the skip-path computation, while the
runtime also writes the layer's own KV (Invariant~1).
The example selects a static layer subset, following the unified
layer-skipping strategy~\citep{liu2024unified}, and reuses the released
checkpoint's projector.
Place the definition in the skipper module,
which already imports the adapter types, registration function and
PyTorch.

\begin{lstlisting}[style=vpy]
class StaticDepthSkipper(FullGraphSkipperAdapter):
    name = "static_depth"
    supported_actions = frozenset((LogicalAction.RUN,
                                   LogicalAction.PROJECT_ONLY))
    requires_router_weights = False
    projector_kind = FLEXIDEPTH_PROJECTOR
    payload_semantics = "binary_static_depth"

    def __init__(self, ratio):
        self.ratio = float(ratio)
        if not 0.0 <= self.ratio <= 1.0:
            raise ValueError("ratio must be in [0, 1]")

    def prepare_batch(self, *, hidden_states, valid_rows,
                      route_layer_order, **_):
        # Choose a tail within the engine's existing routed set.
        take = int(len(route_layer_order) * self.ratio)
        layers = frozenset(
            route_layer_order[len(route_layer_order) - take:])
        project_mask = (~valid_rows).view(-1, 1)
        run_mask = torch.ones_like(project_mask)
        # Preallocate device payloads; padding rows always RUN.
        dtype = hidden_states.dtype
        return (layers, run_mask, run_mask.to(dtype),
                project_mask, project_mask.to(dtype))

    def route(self, hidden_states, *, router, forced_action=None,
              layer_id=None, batch_state=None):
        if forced_action is not None:
            raise ValueError("forced routes are unsupported")
        if layer_id is None or batch_state is None:
            raise ValueError("layer and prepared batch required")
        layers, run, run_w, project, project_w = batch_state
        mask, weights = ((project, project_w) if layer_id in layers
                         else (run, run_w))
        return FullGraphActionBatch(
            adapter_name=self.name,
            supported_actions=self.supported_actions,
            branch_weights=weights, explicit_run_mask=mask,
            payload_semantics=self.payload_semantics)

    def attestation(self):
        return {"name": self.name, "decision": "static_tail_fraction",
                "static_depth_ratio": self.ratio,
                "requires_router_weights": self.requires_router_weights,
                "projector_kind": self.projector_kind,
                "supported_actions": ["run", "project_only"]}

def build_static_depth(arm):
    return StaticDepthSkipper(arm["static_depth_ratio"])

register_skipper("static_depth", build_static_depth)
\end{lstlisting}

\noindent Register the serving configuration in the design module:

\begin{lstlisting}[style=vpy]
ARMS["integrated_staticdepth"] = {
    "skipper": "static_depth", "phases": "both",
    "regime_switch": True, "static_depth_ratio": 0.5,
}
\end{lstlisting}

\noindent The explicit mask and branch weights have shape
\texttt{[rows, 1]}; no threshold or forced action accompanies the mask.
At ratio zero every routed layer runs; at ratio one every valid row
projects throughout the routed set.

The example reuses FlexiDepth's projector execution.
To integrate a different projector, implement its model-loading and
executor paths, register its metadata and checkpoint tensors, and
select it from the policy.

\section{Route-Guided Cohort Execution at the Kernel Level}
\label{app:compaction}
Algorithm~\ref{alg:compaction} gives one routed decode layer.
There are $B$ resident rows in a graph captured for batch size
$C \ge B$; $C$ is the graph's capacity.
The released router supplies a weight $w_i \in [0,1]$ and selects
RUN when $w_i > 0.5$; Qwen uses a straight-through hard gate.
Exact device-side counts determine the cohort sizes, without a
capacity fraction or host synchronization.

Prefill uses the same pack and scatter with cuBLAS over the packed rows.
Captured decode uses count-bounded Triton GEMMs with a device-tuned tile table (Appendix~\ref{app:h100}).
Their accumulation order differs from cuBLAS;
Table~\ref{tab:faithfulness-2x2} evaluates end-task quality through
the routed execution path.

\section{Validation Gates}
\label{app:gates}

\noindent\textbf{The quality comparison.} For each task the paired
per-document difference-in-differences $(D-C)-(B-A)$ has a two-sided $95\%$ $t$ interval, reported as is.

\noindent\textbf{Attestation.} The runtime \emph{attests} what actually
executed through its info endpoint: route and cohort counters, passes
per mode, and every switch decision. The gates below read this
attestation rather than the launch configuration, so a misconfigured
mechanism cannot pass silently.

\noindent\textbf{Serving gates.} Each is enforced in code and can refuse a cell. Grouped by what they protect:
\begin{itemize}
\item \emph{Identity of what is served.} Served design: the design read from the info
  endpoint after boot equals the intended arm's, field by field.
\item \emph{Comparability of the two arms.} Execution difference: the
  two servers' reported execution (graph capture sizes, KV capacity,
  chunk size, backends) matches a declared configuration. Cross-arm identity:
  resolved configurations may differ only in an allow-listed set.
  This covers the runtime's own settings and the KV budget consumed by its
  weights: $3{,}840$ tokens, or $480$~MB at $128$~KB per Llama-3-8B token.
  That is $0.98\%$ of the A100's $393{,}319$-token pool and
  $2.07\%$ of the RTX A6000's $185{,}766$.
  The gate refuses a larger KV pool on the served arm.
\item \emph{Validity of a cell.} Work identity: both arms generated
  the same number of tokens per request, from raw server-reported
  output ids. Accounting: submitted $=$ started $=$ completed $=$
  reported, zero errors, no filtered tail. Declared cell size: the
  exact suite, duration derived as requests over rate. Zero-empty: no
  empty generation in a natural-lane or quality cell, except the cases of Appendix~\ref{app:faithful-eos}.
  Protocol identity: served arms receive the same token ids as the
  PyTorch arms.
\item \emph{Mechanism evidence.} Skipping executed: the attestation
  read before and after the cell must show routed passes on the phases
  the arm declares, with per-mode prefill counters summing to the
  scheduler's pass count. A cell whose declared mechanism never executed is refused.
\end{itemize}
Every arm is a named entry in one module, and an undeclared
difference in the arms' executed settings fails the cell.

\section{The Switching Thresholds from the Roofline}
\label{app:roofline}

\noindent\textbf{Prefill floor.} A coarse sweep over 1k--3k-token
prefill passes put routed mode's crossover near $1.5$k tokens, so
$1{,}536$ was fixed as the floor; it is applied to an admission
round's prompt tokens, before prefix-cache hits.
Like a chunked-prefill size~\citep{agrawal2023sarathi}, this setting
is applied unchanged to every workload, rate, arm and device.

\noindent\textbf{Decode rule.} For a weight-stationary decode GEMM over
$M$ rows with fp16 weights, FLOPs are $2MKN$ and weight bytes $2KN$.
The arithmetic intensity, the FLOPs per byte moved, is therefore $M$.
The device's ridge point becomes a row count,
$M^* = \text{peak FLOP/s} / \text{bandwidth} \approx 161$ on
the A100.
Below $M^*$ the
weights are streamed whether or not a row skips. What a skipped row
saves at a routed layer is that layer's attention read of its own
context, $b = 4$\,KB per resident token for Llama-3-8B. Over $L_r = 16$
routed layers with a fraction $s$ of decisions projecting, a step
holding $V$ resident tokens removes $s\,L_r\,b\,V$ bytes, i.e.
$s\,L_r\,b\,V / \text{BW}$ of time, against routed mode's fixed
per-step cost $\tau = \vpVstarTauMsLSixteen$\,ms.
This cost is the measured intercept difference of the two modes'
per-step latency lines on the A100's 16 routed layers. It is calibrated
once and reused as a per-routed-layer coefficient on every device and
model (scaled by $L_r/16$ for Qwen); only the device bandwidth and the
model and policy constants change. The H100, A6000 and Qwen results
therefore test the rule's transfer without per-device retuning.
Routed mode is profitable when
\[
  V \;>\; V^* = \frac{\tau \cdot \text{BW}}{s\,L_r\,b}
  \;=\; \frac{\vpVstarTauMsLSixteen\,\text{ms} \times \vpVstarPeakTbps\,\text{TB/s}}{0.5 \times 16 \times 4\,\text{KB}}
  \;\approx\; 158\text{k resident tokens},
\]
with the checkpoint's declared decode skip ratio $s = 0.5$ (the
always-route arm attests $\vpLlamaKneeSDecode$ at the GSM8K knee). The
rule estimates the break-even point using the unweighted decision
ratio as a proxy for its context-weighted counterpart.
Above $M^*$, skipped rows also remove compute, making the estimate
conservative.

\noindent\textbf{The served thresholds and the rule.} The A100 thresholds (exit
$160$k, enter $200$k) were measured with a two-mode row sweep at $1$k context. Routed mode ran $7\%$ slower per
step at $128$ rows ($131$k tokens) and $4\%$ faster at $256$ rows
($262$k).
Linear interpolation places the crossover near $214$k; the enter
threshold was set just below it and the exit $20\%$ lower. The rule gives $158$k, within $2\%$ of the exit. Rows alone would misplace the thresholds, since the saving scales with rows $\times$ context.
For another device or model, the rule gives predicted thresholds: exit at $V^*$, enter at $1.25\,V^*$.
This sets the H100, A6000 and Qwen thresholds
(Appendices~\ref{app:h100} and~\ref{app:qwen}). From when $V$ crosses the enter threshold until it falls to the exit threshold, requests that start decoding are pinned to routed mode, and the crossing promotes the dense-decoding requests already running.

\IfFileExists{figures/sweep_prediction.pdf}{%
\begin{figure}[h]
  \centering
  \includegraphics[width=0.62\linewidth]{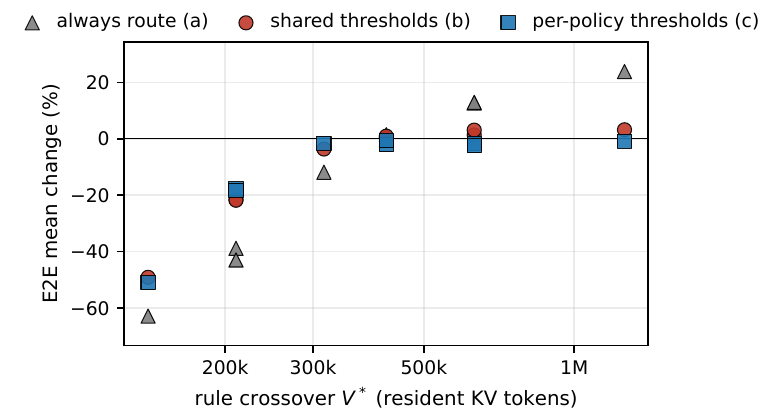}
    \caption{The rule's crossover $V^*$ per RandomSkip arm against its
  measured mean E2E change under the three switch settings of Figure~\ref{fig:sweep}.}
  \label{fig:pred}
\end{figure}}{}
\noindent\textbf{The two rules on this grid.} The grid's skip ratios
are $s = r\,d \in \{\tfrac{1}{16}, \tfrac18, \tfrac{3}{16}, \tfrac14,
\tfrac38, \tfrac{9}{16}\}$. At upstream's estimated knee occupancy of
$\vpPredUpstreamOcc$k tokens the decode rule splits the grid between
$s = \tfrac{3}{16}$ and $s = \tfrac14$; the prefill share threshold sits at $0.35$. One point, $50\%\times50\%$ ($s = \tfrac14$), lies between them and routes only in decode ($\vpSweepEtoEMeanRFiftyDFifty\%$; Figure~\ref{fig:pred}). The
three winners are the arms whose prefill passes are routed under the thresholds: their TTFT moves by $\vpSweepTtftMeanRSeventyfiveDFifty$ to
$\vpSweepTtftMeanBest\%$, the others' by $\vpSweepTtftMeanRFiftyDFifty$
to $\vpSweepTtftMeanWorst\%$. With always route (Figure~\ref{fig:sweep}a) the change runs from
$\vpSweepUngEtoEMeanWorst\%$ at $25\%\times25\%$ to
$\vpSweepUngEtoEMeanBest\%$ at $75\%\times75\%$.
This is the policy's return under routed mode.
Switching rules reduce the cost of unprofitable policies
(Table~\ref{tab:sweep-cells}).

\IfFileExists{generated/sweep_cells_rows.tex}{%
\begin{table}[h]
  \centering
  \caption{The nine RandomSkip arms ($n{=}1$ each): the rule's
  crossover $V^*$ and own thresholds, and per switch setting the cell's
  estimated p90 occupancy (running requests $\times$ mean context), the
  share of decode passes and rows served in routed mode, and the mean
  E2E change (\%).}
  \label{tab:sweep-cells}
  \scriptsize
  \setlength{\tabcolsep}{4pt}
  \begin{tabular}{rrrrrrrrrrr}
  \toprule
  & & & & always route (Fig.~\ref{fig:sweep}a) & \multicolumn{3}{c}{shared thresholds (Fig.~\ref{fig:sweep}b)} & \multicolumn{3}{c}{per-policy thresholds (Fig.~\ref{fig:sweep}c)} \\
  $r$ & $d$ & $V^*$ & own thresholds & E2E $\Delta$ & occupancy & routed & E2E $\Delta$ & occupancy & routed & E2E $\Delta$ \\
  \midrule
  25\,\% & 25\,\% & 1261k & 1260k/1580k & +23.8 & 408k & 95/91\% & +3.2 & 407k & 0/0\% & -0.9 \\
25\,\% & 50\,\% & 631k & 630k/790k & +12.6 & 408k & 95/91\% & +1.3 & 408k & 0/0\% & -1.6 \\
25\,\% & 75\,\% & 420k & 420k/530k & +1.4 & 408k & 95/91\% & +0.2 & 407k & 0/0\% & -2.1 \\
50\,\% & 25\,\% & 631k & 630k/790k & +12.9 & 407k & 95/91\% & +3.1 & 408k & 0/0\% & -2.5 \\
50\,\% & 50\,\% & 315k & 320k/390k & -11.9 & 407k & 95/91\% & -3.6 & 408k & 92/76\% & -1.7 \\
50\,\% & 75\,\% & 210k & 210k/260k & -38.8 & 404k & 93/88\% & -21.5 & 407k & 93/87\% & -17.7 \\
75\,\% & 25\,\% & 420k & 420k/530k & -0.3 & 407k & 95/91\% & +0.9 & 408k & 0/0\% & -0.5 \\
75\,\% & 50\,\% & 210k & 210k/260k & -42.9 & 405k & 93/87\% & -21.8 & 406k & 93/88\% & -18.3 \\
75\,\% & 75\,\% & 140k & 140k/180k & -62.9 & 263k & 93/87\% & -49.1 & 273k & 93/87\% & -51.0 \\
 
  \bottomrule
  \end{tabular}
\end{table}}{}

\section{Faithfulness at the Token Level}
\label{app:faithful-eos}

\subsection{First-Token Agreement}
\label{app:eos-empties}
Routed mode reproduces the checkpoint's first-token choice on
both tokenized inputs in Table~\ref{tab:faithful-eos}.
They represent CoQA \texttt{lm-eval} document~80 and serving request
\texttt{coqa:train:856:15}.
We compare the top-5 first-token log-probabilities under native
PyTorch, upstream SGLang and routed mode.

\begin{table}[h]
  \centering\scriptsize
  \caption{First generated token and its log-probability on the two prompts whose generations are empty. Each served arm matches its PyTorch reference
  within $0.01$ nats.}
  \label{tab:faithful-eos}
  \setlength{\tabcolsep}{4pt}
  \begin{tabular}{lllll}
  \toprule
  prompt & base, PyTorch (A) & upstream SGLang (C) & checkpoint, own code (B) & \sys{} routed (D) \\
  \midrule
  doc 80 & \texttt{Island} $-0.489$ & \texttt{Island} $-0.489$ & \texttt{<eot>} $-0.882$ & \texttt{<eot>} $-0.872$ \\
  856:15 & \texttt{Le} $-0.047$ & \texttt{Le} $-0.047$ & \texttt{<eot>} $-1.033$ & \texttt{<eot>} $-1.033$ \\
  \bottomrule
  \end{tabular}
\end{table}

The checkpoint itself selects \texttt{<|eot\_id|>} on both prompts.
Each served arm matches its native reference within $0.01$ nats:
the immediate stop reflects the checkpoint's first choice.

All four quality arms receive identical token ids.
This controls for a tokenization difference: \texttt{lm-eval}'s HF backend
omits BOS, while the server's text path prepends it.
On doc~80, adding BOS changes the checkpoint's first choice from
\texttt{Island} to \texttt{<eot>}.
The serving gate records \texttt{coqa:train:856:15} as one immediate-stop exception.

\subsection{GSM8K Answer Extraction}
\label{app:extraction}
\texttt{lm-eval} ships two GSM8K filters, and on this checkpoint they
disagree by $11$~pp: flexible extraction puts the checkpoint's cost at
$\vpQGsmBminusAFlex$~pp, strict-match at $\vpQGsmBminusAStrict$. Each
fails on a different arm.

Flexible extraction takes the \emph{last}
number in the generation, and the checkpoint appends a confidence
epilogue after its answer marker
(\texttt{\#\#\#\# 594\textbackslash{}nConfidence: 95\%}), so the filter
reads $95$. The epilogue is the checkpoint's: under its own code the checkpoint emits it on $\vpQGsmArmBFlexFooledPct\%$ of held-out documents, and on
$\vpQGsmArmDFlexFooledPct\%$ when served by \sys{}, while the base model never does.

Strict-match matches
\texttt{\#\#\#\# <number>} and returns \texttt{[invalid]} when the
marker is missing or followed by anything else, such as
\texttt{\#\#\#\# \$21}; the base model under a chat template trips it
on a fifth of its documents. On the $1{,}319$ held-out documents:

\begin{center}\scriptsize
\begin{tabular}{lrr}
\toprule
arm & strict-match unusable & flexible extraction misreads \\
\midrule
A~~base model, \texttt{lm-eval} backend      & $\vpQGsmArmAStrictUnparsed$ ($\vpQGsmArmAStrictUnparsedPct\%$) & $\vpQGsmArmAFlexFooled$ ($\vpQGsmArmAFlexFooledPct\%$) \\
C~~base model, upstream SGLang      & $\vpQGsmArmCStrictUnparsed$ ($\vpQGsmArmCStrictUnparsedPct\%$) & $\vpQGsmArmCFlexFooled$ ($\vpQGsmArmCFlexFooledPct\%$) \\
B~~checkpoint, its own code         & $\vpQGsmArmBStrictUnparsed$ ($\vpQGsmArmBStrictUnparsedPct\%$)  & $\vpQGsmArmBFlexFooled$ ($\vpQGsmArmBFlexFooledPct\%$) \\
D~~checkpoint, served by \sys{}     & $\vpQGsmArmDStrictUnparsed$ ($\vpQGsmArmDStrictUnparsedPct\%$)  & $\vpQGsmArmDFlexFooled$ ($\vpQGsmArmDFlexFooledPct\%$) \\
\bottomrule
\end{tabular}
\end{center}

\noindent The \emph{composite} applies one rule to every arm:
use strict-match when it parses, otherwise flexible extraction.
Both filters come from \texttt{lm-eval}.
The rule preserves the two base-model scores and correctly reads the
checkpoint's confidence epilogues.
Its checkpoint cost is $\vpQGsmBminusA$~pp, between the two
single-filter values.

The two diagnosed format failures are disjoint: strict-match reads
the marked answers whose epilogues mislead flexible extraction.
Disagreements where strict-match parses incorrectly and flexible
extraction is correct affect $0.00\%$ of upstream documents at every rate, at most $0.53\%$ of hybrid documents and $0.61\%$ of always-route documents.
The single-filter comparisons give DiD values of
$\vpQGsmDodFlex \pm \vpQGsmDodFlexCi$~pp with flexible extraction and
$\vpQGsmDodStrict \pm \vpQGsmDodStrictCi$ with strict-match.
Both agree within their intervals with the composite result,
$\vpQGsmDod \pm \vpQGsmDodCi$ in Table~\ref{tab:faithfulness-2x2}.

\subsection{The Served Configuration Across Loads}
\label{app:loaded-sweep}
Table~\ref{tab:loaded-sweep} scores the three served systems at every
rate of the load grid; Table~\ref{tab:faithfulness-2x2} (right) is its knee
row. Below the knee,
where these cells keep GSM8K and CoQA decode in dense mode, the hybrid shows no resolved difference
from upstream ($\vpLqGsmLowDeltaUp \pm \vpLqGsmLowDeltaUpCi$~pp on GSM8K).
At and above the knee, where it routes $\vpLqGsmMidHybRoutedShare$ and
$\vpLqGsmHighHybRoutedShare\%$ of GSM8K decode rows, it shows none from
always-route. On BBH, the hybrid serves $\vpLqBbhMidHybRoutedShare\%$ of decode rows through routed mode at the knee.
Every cell is one repetition with a per-document interval.

\begin{table}[h]
  \centering
  \caption{Task quality of the three served systems across the load
  grid, one repetition per cell, same metrics as
  Table~\ref{tab:faithfulness-2x2}. ``Routed'' is the share of decode rows served in routed mode (\sys{} / always-route).}
  \label{tab:loaded-sweep}
  \scriptsize
  \setlength{\tabcolsep}{4pt}
  \begin{tabular}{llrrrrrr}
  \toprule
  Task & Load & upstream & \sys{} & always-route & routed (\%) & vs.\ upstream (pp) & vs.\ always-route (pp) \\
  \midrule
  \IfFileExists{generated/loaded_quality_sweep.tex}{GSM8K & $0.75\times Q^*$ & 77.63 & 77.26 & 74.00 & 0 / 100 & $-0.38 \pm 1.39$ & $+3.26 \pm 2.34$ \\
GSM8K & $0.95\times Q^*$ & 77.63 & 74.00 & 74.07 & 90 / 100 & $-3.64 \pm 2.22$ & $-0.08 \pm 1.46$ \\
GSM8K & $1.25\times Q^*$ & 77.79 & 74.00 & 73.92 & 99 / 100 & $-3.79 \pm 2.30$ & $+0.08 \pm 0.65$ \\
CoQA & $0.75\times Q^*$ & 76.74 & 76.74 & 77.66 & 0 / 100 & $+0.00 \pm 0.00$ & $-0.92 \pm 1.94$ \\
CoQA & $0.95\times Q^*$ & 76.74 & 77.68 & 77.73 & 71 / 100 & $+0.94 \pm 0.82$ & $-0.05 \pm 1.76$ \\
CoQA & $1.25\times Q^*$ & 76.71 & 79.65 & 77.66 & 99 / 100 & $+2.94 \pm 1.55$ & $+1.99 \pm 1.45$ \\
BBH & $0.75\times Q^*$ & 67.93 & 59.35 & 59.58 & 98 / 100 & $-8.59 \pm 1.06$ & $-0.23 \pm 0.52$ \\
BBH & $0.95\times Q^*$ & 67.87 & 59.47 & 59.68 & 99 / 100 & $-8.40 \pm 1.07$ & $-0.22 \pm 0.38$ \\
BBH & $1.25\times Q^*$ & 67.98 & 59.45 & 59.59 & 100 / 100 & $-8.52 \pm 1.07$ & $-0.14 \pm 0.26$ \\
 }{}
  \bottomrule
  \end{tabular}
\end{table}

\section{Latency and Makespan Across Load}
\label{app:map}
The benchmark is open-loop with fixed work and a common injection
window. Each arm's makespan includes that window and its final drain.
While the server keeps up with arrivals, execution savings appear
as lower latency: smaller concurrent batches, shorter token intervals
and less queueing. Makespan moves only through the drain, the last and
longest requests to finish, so it reflects the tail of each cell (Figure~\ref{fig:map}).
\IfFileExists{figures/latency_map.pdf}{%
\begin{figure}[h]
  \centering
  \includegraphics[width=\linewidth]{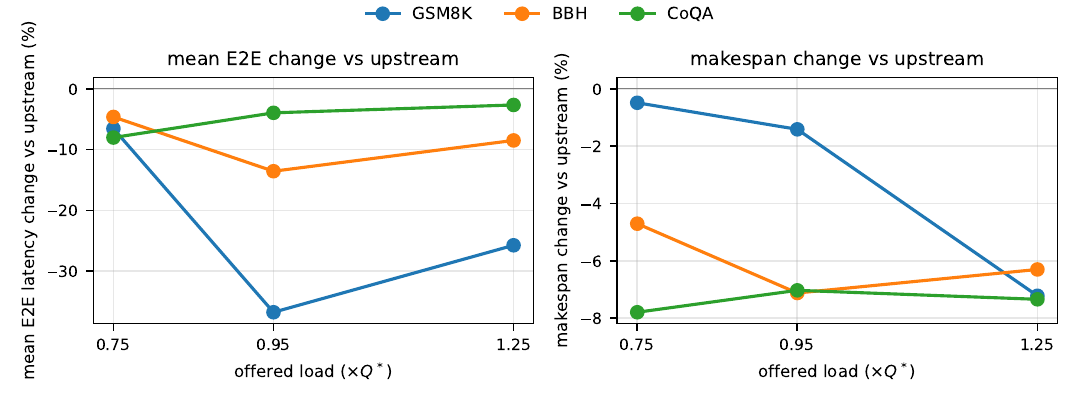}
  \caption{Mean E2E change (left) and makespan change (right) against
  offered load, per workload, from the cells of
  Table~\ref{tab:v13-headline}.}
  \label{fig:map}
\end{figure}

The E2E reduction is roughly uniform across output lengths at and
below the knee on GSM8K and BBH.
Binning by pinned output tokens gives per-bin spreads of $\vpMapSpreadGsmLow$ and
$\vpMapSpreadGsmMid$~pp on GSM8K and $\vpMapSpreadBbhLow$ and
$\vpMapSpreadBbhMid$~pp on BBH. It widens to $\vpMapSpreadGsmHigh$~pp
on GSM8K in overload and to
$\vpMapSpreadCoqaHigh$--$\vpMapSpreadCoqaLow$~pp on CoQA, where a
request's place in the queue sets its gain.
}{}

Mean TTFT falls on GSM8K at every rate, where prefill routing shortens
the prefill passes (Table~\ref{tab:v13-headline}). On BBH and CoQA it shows
no resolved difference from upstream: prefill routes few passes on CoQA
(Table~\ref{tab:engagement}), and on BBH prefill-only routing stays near
upstream (Table~\ref{tab:ablation}).

\noindent\textbf{Throughput under saturation.} Table~\ref{tab:capacity}
counts the output tokens emitted and the requests completed between the
first and the last arrival of each $1.25\times Q^*$ cell, the injection
window both arms share; the drain is excluded. Both arms are saturated
there: in every repetition of every column they complete only
$\vpCapDoneArrivedMin$--$\vpCapDoneArrivedMax$ of the requests that
arrive inside the window, and their backlog grows through it. On GSM8K
and CoQA the running batch stays full while the waiting queue grows, and
TPS and RPS agree within one percentage point. On BBH, whose shared few-shot prefixes
keep each request cheap in KV, the backlog accumulates in the running
batch; in-flight requests grow through the window, and the RPS change
($\vpCapBbhRps\%$) exceeds the TPS change ($\vpCapBbhTps\%$).

\section{Natural Stopping and the Output Lengths}
\label{app:banks}
The matched-work lane replays output lengths recorded from the base
model's natural generation.
Here the base model on upstream SGLang and the FlexiDepth checkpoint
on \sys{} each generate to their own stop.
Both receive the same open-loop arrivals, with one repetition per cell
(Table~\ref{tab:natural-stacks}).
A generation reaching its remaining context window is counted as a
context-limit hit.

\begin{table}[h]
  \centering
  \caption{The two stacks under natural stopping, one repetition per
  cell: mean output tokens per request, context-limit hits, mean E2E
  latency and makespan. A third row serves the checkpoint in the
  always-route configuration.}
  \label{tab:natural-stacks}
  \scriptsize
  \setlength{\tabcolsep}{4pt}
  \begin{tabular}{llrrrrr}
  \toprule
  Workload & Load ($\times Q^*$) & Stack & out tokens & limit hits & E2E s & makespan s \\
  \midrule
  \IfFileExists{generated/natural_lane_rows.tex}{GSM8K & 0.75 & Base + upstream & 116 & 3 & 6.1 & 437 \\
 &  & FlexiDepth + \sys{} & 126 & 5 & 6.4 & 440 \\
 & & FlexiDepth always-route & 299 & 82 & 18.8 & 632 \\
GSM8K & 0.95 & Base + upstream & 116 & 3 & 22.4 & 388 \\
 &  & FlexiDepth + \sys{} & 330 & 99 & 38.2 & 668 \\
 & & FlexiDepth always-route & 302 & 84 & 32.4 & 612 \\
GSM8K & 1.25 & Base + upstream & 116 & 3 & 57.2 & 388 \\
 &  & FlexiDepth + \sys{} & 271 & 69 & 61.6 & 563 \\
 & & FlexiDepth always-route & 297 & 81 & 64.6 & 598 \\
\addlinespace
BBH & 0.75 & Base + upstream & 257 & 42 & 14.0 & 349 \\
 &  & FlexiDepth + \sys{} & 394 & 114 & 24.4 & 548 \\
 & & FlexiDepth always-route & 408 & 121 & 25.8 & 565 \\
BBH & 0.95 & Base + upstream & 255 & 41 & 21.9 & 329 \\
 &  & FlexiDepth + \sys{} & 405 & 120 & 31.9 & 551 \\
 & & FlexiDepth always-route & 420 & 128 & 33.8 & 576 \\
BBH & 1.25 & Base + upstream & 262 & 45 & 36.3 & 336 \\
 &  & FlexiDepth + \sys{} & 405 & 121 & 44.7 & 545 \\
 & & FlexiDepth always-route & 413 & 125 & 46.0 & 565 \\
\addlinespace
CoQA & 0.75 & Base + upstream & 166 & 82 & 8.9 & 563 \\
 &  & FlexiDepth + \sys{} & 156 & 76 & 7.5 & 497 \\
 & & FlexiDepth always-route & 83 & 38 & 3.9 & 393 \\
CoQA & 0.95 & Base + upstream & 164 & 81 & 11.9 & 546 \\
 &  & FlexiDepth + \sys{} & 125 & 59 & 8.8 & 410 \\
 & & FlexiDepth always-route & 86 & 40 & 7.3 & 362 \\
CoQA & 1.25 & Base + upstream & 166 & 82 & 25.1 & 541 \\
 &  & FlexiDepth + \sys{} & 161 & 78 & 24.0 & 478 \\
 & & FlexiDepth always-route & 84 & 39 & 20.4 & 348 \\
 }{}
  \bottomrule
  \end{tabular}
\end{table}

Matched work measures the runtime; natural stopping characterizes each
model--runtime stack, whose generated work differs.
At the GSM8K knee the checkpoint generates $\vpNatGsmMidOutTokVs$
output tokens per request against the base model's
$\vpNatGsmMidOutTokUp$ ($\vpNatGsmMidOutTok\%$).
It reaches the context limit on $\vpNatGsmMidCapVs$ requests against
$\vpNatGsmMidCapUp$; on BBH it generates
$\vpNatBbhLowOutTokAbs$--$\vpNatBbhMidOutTokAbs\%$ more.
The long requests carry most of a cell's tokens, so each stack's latency
and makespan follow its generated work. They rise on GSM8K and BBH.
On CoQA the checkpoint generates similar or less work
($\vpNatCoqaHighOutTok$ to $\vpNatCoqaMidOutTok\%$ output tokens), and its
stack finishes sooner (E2E $\vpNatCoqaHighEtoE$ to $\vpNatCoqaMidEtoE\%$).
These are stack-level differences, not runtime effects.

The matched-work cells freeze the base model's per-request lengths,
separating the serving engine's execution cost from changes in generated
work. In this natural lane, requests run to EOS or their remaining
context window ($\vpSuiteRemainingCtxMin$--$\vpSuiteRemainingCtxMax$
tokens), with context-limit hits counted explicitly.
Task quality is measured separately under \texttt{lm-eval}'s
per-task generation limits.

\subsection{Native Checkpoint Repetition Analysis}
\label{app:attribution}
Under the checkpoint authors' native PyTorch execution, FlexiDepth
increases the incidence of runaway generation over its base model on
the same prompts (Table~\ref{tab:attribution}).
Both the base model and checkpoint generate greedily from frozen
token ids within an 8192-token window.
On a seeded random sample
of $\vpAttrFdPopN$ prompts of the GSM8K knee cell, the base model repeats on none and the checkpoint on $\vpAttrFdPopPct \pm \vpAttrFdPopCi\%$,
each a degenerate repetition of the answer line.

The propensity is prompt-specific.
The checkpoint repeats on $\vpAttrFdBOneVskPct\%$ of the $\vpAttrRawVskN$ served-arm repeaters, against
$\vpAttrFdBOneCtlPct\%$ of the controls.
The sampled groups stop once repetition is detected.

Under serving, which prompts repeat varies with load and stack
(Table~\ref{tab:natural-stacks}). Sampling penalties, such as repetition and frequency penalties, are a
standard mitigation. The evaluation keeps greedy decoding under the
third-party \texttt{lm-eval} protocol, and tuning these parameters is left
to future work.

\begin{table}[h]
  \centering
  \caption{Native repetition attribution on the GSM8K knee cell's prompts
  (PyTorch, base model and checkpoint). A repetition is a degenerate
  repeated answer or a generation that reached the window (cap).}
  \label{tab:attribution}
  \scriptsize
  \setlength{\tabcolsep}{3pt}
  \begin{tabular}{llrrrrrr}
  \toprule
  Model & Prompt group & prompts & repeats & \% & cap & mean & p90 \\
  \midrule
  \IfFileExists{generated/attribution_rows.tex}{    Base, batch 16 & random sample & 500 & 0 & 0.0 & 0 & 108 & 163 \\
     & of which served-arm repeaters & 15 & 0 & 0.0 & 0 & 100 & 153 \\
     & of which never repeated & 485 & 0 & 0.0 & 0 & 109 & 164 \\
    \midrule
    FlexiDepth, batch 16 & random sample & 500 & 11 & 2.2 & 0 & 155 & 213 \\
     & of which served-arm repeaters & 15 & 4 & 26.7 & 0 & 364 & 1024 \\
     & of which never repeated & 485 & 7 & 1.4 & 0 & 148 & 209 \\
    \midrule
    Base, batch 1 & served-arm repeaters & 143 & 0 & 0.0 & 0 & 107 & 153 \\
     & upstream repeaters & 3 & 2 & 66.7 & 2 & 4656 & 6956 \\
     & controls & 100 & 0 & 0.0 & 0 & 102 & 145 \\
    \midrule
    FlexiDepth, batch 1 & served-arm repeaters & 143 & 29 & 20.3 & 29 & 1583 & 7278 \\
     & upstream repeaters & 3 & 0 & 0.0 & 0 & 194 & 215 \\
     & controls & 100 & 2 & 2.0 & 2 & 273 & 196 \\
    \midrule
    FlexiDepth, batch 16 & served-arm repeaters & 143 & 29 & 20.3 & 29 & 1518 & 7009 \\
     & upstream repeaters & 3 & 0 & 0.0 & 0 & 188 & 222 \\
     & controls & 100 & 3 & 3.0 & 3 & 330 & 196 \\
 }{}
  \bottomrule
  \end{tabular}
\end{table}

\section{Mechanism Ablation}
\label{app:ablation}
\newcommand{\vpAblBbhBlock}{}
\newcommand{\vpAblCoqaBlock}{}
\IfFileExists{generated/ablation_rows_bbh_cot.tex}{\renewcommand{\vpAblBbhBlock}{\midrule \multicolumn{11}{l}{\textit{BBH}} \\ 0.75$\times$ & -4.6 $\pm$ 0.5 & -4.7 $\pm$ 0.5 & -3.7 & -4.7 & -0.3 & +0.1 & -3.2 & -5.7 & +2.0 & -2.8 \\
0.95$\times$ & -13.6 $\pm$ 1.2 & -7.1 $\pm$ 0.2 & -16.0 & -6.2 & +0.5 & +0.1 & -13.0 & -6.7 & -9.9 & -3.8 \\
1.25$\times$ & -8.5 $\pm$ 1.0 & -6.3 $\pm$ 0.7 & -9.7 & -5.9 & -0.2 & +0.1 & -9.0 & -6.1 & -6.3 & -3.3 \\
 }}{}
\IfFileExists{generated/ablation_rows_coqa.tex}{\renewcommand{\vpAblCoqaBlock}{\midrule \multicolumn{11}{l}{\textit{CoQA}} \\ 0.75$\times$ & -8.0 $\pm$ 0.6 & -7.8 $\pm$ 0.3 & -8.0 & -7.8 & +0.8 & +0.4 & -1.3 & -5.1 & +3.4 & +0.5 \\
0.95$\times$ & -4.0 $\pm$ 4.4 & -7.0 $\pm$ 0.4 & -2.6 & -7.3 & -0.9 & +0.3 & +5.5 & -6.1 & +4.9 & +5.3 \\
1.25$\times$ & -2.7 $\pm$ 0.2 & -7.3 $\pm$ 0.1 & -2.7 & -8.1 & +0.1 & +0.1 & +0.0 & -7.1 & -1.1 & -4.0 \\
 }}{}
\IfFileExists{generated/ablation_rows.tex}{%
\begin{table}[h]
  \centering
  \caption{Mechanism ablation: mean E2E and makespan change (\%)
  against upstream at the main output lengths. The served arm is the $n{=}6$ main result; each ablation arm is one repetition,
  and negative is better.}
  \label{tab:ablation}
  \scriptsize
  \setlength{\tabcolsep}{3pt}
  \begin{tabular}{lrrrrrrrrrr}
  \toprule
  & \multicolumn{2}{c}{served ($n{=}6$)} & \multicolumn{2}{c}{decode only} & \multicolumn{2}{c}{prefill only} & \multicolumn{2}{c}{always route} & \multicolumn{2}{c}{no promotion} \\
  rate & E2E & makespan & E2E & makespan & E2E & makespan & E2E & makespan & E2E & makespan \\
  \midrule
  \multicolumn{11}{l}{\textit{GSM8K}} \\
  0.75$\times$ & -6.6 $\pm$ 1.5 & -0.5 $\pm$ 0.1 & +0.4 & +0.0 & -8.6 & -0.7 & -5.5 & -1.0 & -6.7 & -0.5 \\
0.95$\times$ & -36.8 $\pm$ 2.0 & -1.4 $\pm$ 0.1 & -14.3 & -0.9 & -35.6 & -1.2 & -46.5 & -1.9 & -32.6 & -1.2 \\
1.25$\times$ & -25.7 $\pm$ 0.7 & -7.2 $\pm$ 0.2 & -7.8 & -2.6 & -19.9 & -5.4 & -26.3 & -7.4 & -24.5 & -7.0 \\
 
  \vpAblBbhBlock
  \vpAblCoqaBlock
  \bottomrule
  \end{tabular}
\end{table}}{}
Four arms isolate the mechanisms of \S\ref{sec:design-production} and \S\ref{sec:design-engagement} on
all three workloads at the three main rates, one repetition each, paired against upstream at the main output lengths (Table~\ref{tab:ablation}). \emph{Decode only}: routed mode on the decode passes the decode thresholds select, prefill dense. \emph{Prefill only}:
routed mode on the prefill rounds the admission rule selects, decode dense. \emph{Always-route}: both phases without the mode switch. \emph{No promotion}: the served hybrid without the one-way
promotion into routed decode (\S\ref{sec:design-engagement}).
All routed arms share the cohort execution path
(Appendix~\ref{app:compaction}).

\noindent\textbf{GSM8K.} Prefill routing provides the gain:
its knee-cell E2E change, $\vpAblVpreBinarycohortEtoEMid\%$, is close
to the served arm's $\vpEtoEMeanGsmMid\%$.
Decode-only remains inactive below the knee.
Always-route is faster at the knee in this single repetition;
the CoQA paragraph shows what the switch gains in exchange.

\noindent\textbf{BBH.} Decode routing provides the gain across the load grid; prefill-only stays near upstream.
The served arm is faster than always-route at and below the knee.
The switch avoids routed-mode costs on passes it keeps dense.

\noindent\textbf{CoQA.} The gain comes from finishing the long-output tail sooner:
the served and decode-only arms improve makespan by $\vpMakespanCoqaMidAbs$--$8.1\%$.
The switch matters most at the knee, where always-route is
$\vpAblCoqaIntegratedAlwaysskipEtoEMidVsHeadAbs$~pp behind the served arm in mean E2E.

\subsection{Kernel Headroom}
\label{app:kernel-headroom}
Routed mode's GEMMs use count-bounded Triton kernels so captured
execution can consume row counts held on the device.
Outside capture, the same rows go through cuBLAS.
The tile tuner times cuBLAS on the same packed rows beside each candidate.
On the A100, the served kernels take a median of
$\vpTileRatioMedAHundred\times$ cuBLAS's time
($\vpTileRatioMinAHundred$--$\vpTileRatioMaxAHundred\times$
over the 20 operation/count-range shapes).
On the RTX A6000, the median is $\vpTileRatioMedAsix\times$
($\vpTileRatioMinAsix$--$\vpTileRatioMaxAsix\times$).
The serving gains already include these count-bounded kernels.
The comparisons identify target shapes for further kernel optimization.

\section{Tail Latencies}
\label{app:tails}
Table~\ref{tab:tails} gives the p50 and p99 companions of Table~\ref{tab:v13-headline}; Figure~\ref{fig:sweep-p99} the p99 companion of the RandomSkip maps.

\IfFileExists{figures/sweep_heatmap_v2_p99.pdf}{%
\begin{figure}[h]
  \centering
  \begin{minipage}[t]{0.30\linewidth}\centering\includegraphics[width=\linewidth]{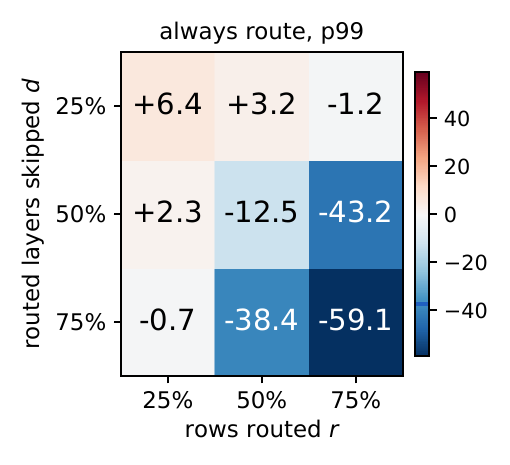}\\[-2pt]{\scriptsize (a) always route}\end{minipage}%
  \hfill%
  \begin{minipage}[t]{0.30\linewidth}\centering\includegraphics[width=\linewidth]{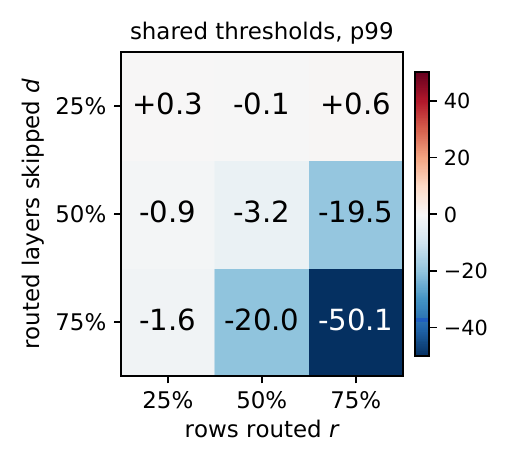}\\[-2pt]{\scriptsize (b) shared thresholds}\end{minipage}%
  \hfill%
  \begin{minipage}[t]{0.30\linewidth}\centering\includegraphics[width=\linewidth]{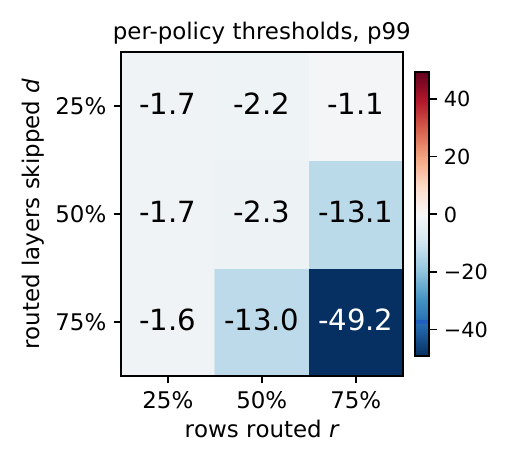}\\[-2pt]{\scriptsize (c) per-policy thresholds}\end{minipage}%
  \caption{p99 E2E change against upstream for the nine RandomSkip arms
  under the three switch settings, the same cells as
  Figure~\ref{fig:sweep}.}
  \label{fig:sweep-p99}
\end{figure}}{}
\begin{table}[h]
  \centering
  \caption{p50 and p99 companions of Table~\ref{tab:v13-headline}: mean
  percentage change with $95\%$ intervals over the same six paired
  repetitions.}
  \label{tab:tails}
  \scriptsize
  \setlength{\tabcolsep}{3pt}
  \begin{tabular}{llrrrrrrr}
  \toprule
  & & & \multicolumn{2}{c}{E2E} & \multicolumn{2}{c}{TPOT} & \multicolumn{2}{c}{TTFT} \\
  \cmidrule(lr){4-5}\cmidrule(lr){6-7}\cmidrule(lr){8-9}
  Workload & Rate & $n$ & p50 & p99 & p50 & p99 & p50 & p99 \\
  \midrule
  \IfFileExists{generated/tails_rows.tex}{BBH & $0.75\times Q^*$ & 6 & -3.0 $\pm$ 0.5 & -8.2 $\pm$ 1.0 & -5.0 $\pm$ 0.8 & -10.7 $\pm$ 2.7 & -0.0 $\pm$ 2.8 & +12.0 $\pm$ 37.5 \\
BBH & $0.95\times Q^*$ & 6 & -15.6 $\pm$ 1.7 & -11.4 $\pm$ 0.3 & -13.8 $\pm$ 0.8 & -22.4 $\pm$ 1.5 & -10.3 $\pm$ 2.4 & +6.8 $\pm$ 31.3 \\
BBH & $1.25\times Q^*$ & 6 & -10.4 $\pm$ 1.1 & -8.9 $\pm$ 0.9 & -10.1 $\pm$ 1.2 & -10.4 $\pm$ 2.0 & +0.7 $\pm$ 11.6 & -16.4 $\pm$ 7.1 \\
CoQA & $0.75\times Q^*$ & 6 & +1.8 $\pm$ 2.6 & -10.1 $\pm$ 0.4 & +2.1 $\pm$ 2.1 & -2.0 $\pm$ 18.4 & -0.6 $\pm$ 2.7 & +5.4 $\pm$ 30.0 \\
CoQA & $0.95\times Q^*$ & 6 & +2.5 $\pm$ 9.9 & -9.2 $\pm$ 0.5 & +4.2 $\pm$ 11.3 & +1.0 $\pm$ 24.0 & -2.6 $\pm$ 7.8 & +0.8 $\pm$ 5.2 \\
CoQA & $1.25\times Q^*$ & 6 & -0.3 $\pm$ 0.4 & -9.1 $\pm$ 0.2 & -2.5 $\pm$ 0.6 & -1.7 $\pm$ 0.6 & +0.3 $\pm$ 0.8 & +0.3 $\pm$ 0.7 \\
GSM8K & $0.75\times Q^*$ & 6 & -6.3 $\pm$ 1.5 & -9.7 $\pm$ 2.8 & -6.3 $\pm$ 1.6 & -10.0 $\pm$ 3.7 & -6.3 $\pm$ 0.6 & -4.4 $\pm$ 4.8 \\
GSM8K & $0.95\times Q^*$ & 6 & -38.4 $\pm$ 1.9 & -37.3 $\pm$ 2.5 & -43.6 $\pm$ 2.1 & -15.3 $\pm$ 3.8 & -53.8 $\pm$ 11.6 & -74.5 $\pm$ 3.0 \\
GSM8K & $1.25\times Q^*$ & 6 & -26.4 $\pm$ 0.7 & -21.2 $\pm$ 0.6 & -10.3 $\pm$ 0.2 & -18.9 $\pm$ 1.7 & -41.8 $\pm$ 1.3 & -40.7 $\pm$ 0.9 \\%
 }{}
  \bottomrule
  \end{tabular}
\end{table}

\section{Absolute Baseline Numbers}
\label{app:absolutes}
Table~\ref{tab:absolutes} gives the upstream arm's absolute values
behind the relative rows of Table~\ref{tab:v13-headline}, averaged over the same repetitions. The artifact separately records the
served arm's absolute means directly. The injected arrival rate matches the offered
rate within $2\%$ on every cell. Makespan exceeds the injection window
by the drain, the tail of the longest pinned requests; on CoQA the
drain is $\vpAbsDrainShareCoqaLow$--$\vpAbsDrainShareCoqaHigh\%$ of the
makespan at every load, so its gain appears in makespan.

\noindent\textbf{Decode-graph coverage.} Both arms capture decode
graphs up to $1{,}024$ rows. The load samples show how often a decode batch exceeded SGLang's default A100 capture setting of $256$ rows, which the default
configuration would have run without a graph. For the baseline it is
$\vpCovGsmMidUpAboveTwoFiftySix\%$ at the GSM8K knee,
$\vpCovBbhMidUpAboveTwoFiftySix\%$ at the BBH knee and
$\vpCovCoqaMidUpAboveTwoFiftySix\%$ at the CoQA knee; for the served
arm, $\vpCovGsmMidVsAboveTwoFiftySix$, $\vpCovBbhMidVsAboveTwoFiftySix$ and
$\vpCovCoqaMidVsAboveTwoFiftySix\%$. Samples exceed
$1{,}024$ rows only at BBH overload, on
$\vpCovBbhHighUpAboveOneKtwentyFour\%$ (baseline) and
$\vpCovBbhHighVsAboveOneKtwentyFour\%$ (served) of samples; there
upstream runs uncaptured and \sys{} replays its captured graphs in capture-sized chunks.

\noindent\textbf{Prefix cache.} The radix prefix cache is enabled on
both arms. The cached share of prefill tokens, upstream\,/\,\sys{},
is $\vpCacheGsm\,/\,\vpCacheGsmVs\%$ on GSM8K,
$\vpCacheBbh\,/\,\vpCacheBbhVs\%$ on BBH (shared few-shot prefixes)
and $\vpCacheCoqa\,/\,\vpCacheCoqaVs\%$ on CoQA, within
$\vpCacheArmDiffMax$~pp between the arms. \sys{} keeps separate dense and
routed namespaces keyed by a request's prefill mode, so a prompt reuses
only KV computed under its own prefill mode, and a request whose decode
mode differs from its prefill mode does not insert its generated KV.

\begin{table}[h]
  \centering
  \caption{Upstream SGLang absolutes per cell, means over the repetitions; drain is makespan minus the injection window.}
  \label{tab:absolutes}
  \scriptsize
  \setlength{\tabcolsep}{4pt}
  \begin{tabular}{llrrrrrrrr}
  \toprule
  Workload & $\times Q^*$ & TTFT (ms) & TPOT (ms) & E2E (s) & makespan (s) & drain (s) & tok/s & req/s injected & req/s offered \\
  \midrule
  \IfFileExists{generated/absolute_rows.tex}{GSM8K & 0.75 & 197 & 52.8 & 6.1 & 437 & 62 & 955 & 9.6 & 9.75 \\
GSM8K & 0.95 & 894 & 238.3 & 26.9 & 390 & 94 & 1087 & 12.2 & 12.35 \\
GSM8K & 1.25 & 29017 & 274.2 & 58.8 & 377 & 152 & 1106 & 16.0 & 16.25 \\
BBH & 0.75 & 132 & 64.4 & 14.4 & 354 & 195 & 2943 & 25.2 & 25.50 \\
BBH & 0.95 & 250 & 118.2 & 23.9 & 330 & 205 & 3111 & 31.9 & 32.30 \\
BBH & 1.25 & 1251 & 200.5 & 38.7 & 323 & 228 & 3180 & 42.0 & 42.50 \\
CoQA & 0.75 & 211 & 138.5 & 8.8 & 561 & 345 & 1169 & 18.5 & 18.75 \\
CoQA & 0.95 & 493 & 796.7 & 13.1 & 543 & 373 & 1193 & 23.5 & 23.75 \\
CoQA & 1.25 & 10204 & 1809.5 & 30.8 & 534 & 405 & 1240 & 30.9 & 31.25 \\
 }{}
  \bottomrule
  \end{tabular}
\end{table}
\begin{table}[h]
  \centering
  \caption{Pass share per phase: the fraction of the served arm's prefill
  and decode passes that ran in routed mode (a routed pass may also
  carry dense requests), from each cell's attestation.
  Means use the six repetition logs.}
  \label{tab:engagement}
  \scriptsize
  \setlength{\tabcolsep}{4pt}
  \begin{tabular}{llrrr}
  \toprule
  Workload & $\times Q^*$ & $n$ & prefill passes routed (\%) & decode passes routed (\%) \\
  \midrule
  \IfFileExists{generated/engagement_rows.tex}{GSM8K & 0.75 & 6 & 24 & 0.0 \\
GSM8K & 0.95 & 6 & 42 & 92.9 \\
GSM8K & 1.25 & 6 & 94 & 99.0 \\
BBH & 0.75 & 6 & 37 & 95.9 \\
BBH & 0.95 & 6 & 54 & 98.7 \\
BBH & 1.25 & 6 & 67 & 99.2 \\
CoQA & 0.75 & 6 & 0 & 69.8 \\
CoQA & 0.95 & 6 & 1 & 97.1 \\
CoQA & 1.25 & 6 & 0 & 99.8 \\
 }{}
  \bottomrule
  \end{tabular}
\end{table}

\section{Two Qwen3 Skippers: the Profitability Boundary}
\label{app:qwen}
Two skippers we trained test where the method starts to be profitable. Both are
alignment-only straight-through-gate routers and projectors on Qwen3
with thinking off, routing layers 18--35 of 36. Each is served by the
same runtime with the thresholds the rule derives from its declared skip
(Appendix~\ref{app:roofline}); training and checkpoint selection are in
Appendix~\ref{app:repro}. We report Qwen3-4B at step $20{,}000$ and
three Qwen3-8B checkpoints. Table~\ref{tab:qwen-serving} identifies each
checkpoint, its decode thresholds and its attested always-route decode skip.
The thresholds use skip-rate estimates from a 32-request training-time routing probe: $0.25$ for 4B, and $\vpProbeSkipQwenEightOld$, $\vpProbeSkipQwenEightAlt$, and $\vpProbeSkipQwenEight$ 
for the three 8B checkpoints. These probe estimates are used only for pre-serving threshold calibration.

Each follows the main matched-work protocol: its own upstream $Q^*$ ($17$ and $14$~req/s), its own output lengths, GSM8K at the three
rates, three paired repetitions, bf16. In both Qwen arms the
Project-Only KV path applies Qwen3's per-head q/k norms and projects
all rows. The Qwen natural lane runs under an
$8{,}192$-token output budget (Qwen3's context window is $40{,}960$
tokens); the quality suites use \texttt{lm-eval}'s per-task limits.

\noindent\textbf{Serving.} Higher skipping translates into larger
serving gains in the three-checkpoint comparison.
The highest-skipping Qwen3-8B checkpoint lowers mean E2E the most at
every load, with each interval excluding zero
(Table~\ref{tab:qwen-serving}).
Qwen3-4B shows no resolved change in mean E2E at any load.
The table also reports its single-repetition run under the shared Llama thresholds.

\begin{table}[h]
  \centering
  \caption{Qwen3 checkpoints against upstream: matched-work
  percentage changes with $95\%$ intervals (A100, bf16, GSM8K;
  $n{=}3$, except the Llama-threshold rows). Group headings give the checkpoint, exit/enter thresholds and always-route decode skip at the knee.}
  \label{tab:qwen-serving}
  \scriptsize
  \setlength{\tabcolsep}{4pt}
  \begin{tabular}{@{}Zlrrrrrrrr@{}}
  \toprule
  & & & \multicolumn{2}{c}{E2E} & \multicolumn{2}{c}{TPOT} & \multicolumn{2}{c}{TTFT} & \\
  \cmidrule(lr){4-5}\cmidrule(lr){6-7}\cmidrule(lr){8-9}
  & Rate & $n$ & mean & p99 & mean & p99 & mean & p99 & makespan \\
  \midrule
  \multicolumn{10}{l}{\textbf{Qwen3-4B, step 20k}; thresholds $320$k/$390$k, skip $\vpQwenFourKneeSDecode$} \\
  \IfFileExists{generated/qwen_serving_rows.tex}{Qwen3-4B & $0.75\times Q^*$ & 3 & +1.2 $\pm$ 1.9 & +1.1 $\pm$ 2.3 & +1.3 $\pm$ 1.9 & +2.7 $\pm$ 1.9 & -0.2 $\pm$ 7.6 & +0.3 $\pm$ 58.9 & +0.1 $\pm$ 0.4 \\
Qwen3-4B & $0.95\times Q^*$ & 3 & +2.7 $\pm$ 2.8 & +0.9 $\pm$ 2.3 & +1.0 $\pm$ 1.1 & +0.8 $\pm$ 0.7 & +9.3 $\pm$ 11.2 & +10.4 $\pm$ 10.2 & +4.5 $\pm$ 0.7 \\
Qwen3-4B & $1.25\times Q^*$ & 3 & +0.6 $\pm$ 1.0 & +0.7 $\pm$ 1.8 & -0.0 $\pm$ 0.6 & -0.2 $\pm$ 2.8 & +1.1 $\pm$ 1.3 & +1.4 $\pm$ 2.4 & +3.1 $\pm$ 0.7 \\%
 }{}
  \addlinespace
  \multicolumn{10}{l}{\textbf{Qwen3-8B, penalty $2{\times}10^{-4}$, step 10k}; thresholds $180$k/$220$k, skip $\vpQwenEightOldKneeSDecode$} \\
  \IfFileExists{generated/qwen8b_old_serving_rows.tex}{Qwen3-8B $2{\times}10^{-4}$ 10k & $0.75\times Q^*$ & 3 & -6.4 $\pm$ 2.1 & -4.5 $\pm$ 2.1 & -2.3 $\pm$ 0.5 & -1.1 $\pm$ 3.1 & -19.5 $\pm$ 4.7 & -18.7 $\pm$ 5.1 & +1.1 $\pm$ 0.2 \\
Qwen3-8B $2{\times}10^{-4}$ 10k & $0.95\times Q^*$ & 3 & -3.8 $\pm$ 1.7 & -3.7 $\pm$ 0.9 & -2.1 $\pm$ 0.2 & -1.1 $\pm$ 6.2 & -4.9 $\pm$ 2.8 & -4.9 $\pm$ 0.9 & +0.6 $\pm$ 0.7 \\
Qwen3-8B $2{\times}10^{-4}$ 10k & $1.25\times Q^*$ & 3 & -3.0 $\pm$ 0.2 & -3.0 $\pm$ 0.7 & -2.3 $\pm$ 0.3 & -1.9 $\pm$ 1.4 & -3.2 $\pm$ 0.2 & -3.3 $\pm$ 0.4 & +0.3 $\pm$ 0.4 \\%
 }{}
  \addlinespace
  \multicolumn{10}{l}{\textbf{Qwen3-8B, penalty $2{\times}10^{-4}$, step 18.75k}; thresholds $160$k/$200$k, skip $\vpQwenEightAltKneeSDecode$} \\
  \IfFileExists{generated/qwen8b_alt_serving_rows.tex}{Qwen3-8B $2{\times}10^{-4}$ 18.75k & $0.75\times Q^*$ & 3 & -6.5 $\pm$ 4.6 & -4.4 $\pm$ 3.7 & -2.3 $\pm$ 1.6 & -1.3 $\pm$ 3.6 & -19.7 $\pm$ 13.8 & -20.9 $\pm$ 8.7 & +0.7 $\pm$ 0.9 \\
Qwen3-8B $2{\times}10^{-4}$ 18.75k & $0.95\times Q^*$ & 3 & -5.0 $\pm$ 2.2 & -4.2 $\pm$ 1.8 & -2.5 $\pm$ 0.7 & -3.0 $\pm$ 4.1 & -6.7 $\pm$ 3.3 & -6.3 $\pm$ 2.4 & -0.2 $\pm$ 1.1 \\
Qwen3-8B $2{\times}10^{-4}$ 18.75k & $1.25\times Q^*$ & 3 & -3.5 $\pm$ 0.9 & -3.3 $\pm$ 0.7 & -2.5 $\pm$ 0.2 & -1.2 $\pm$ 2.3 & -3.9 $\pm$ 1.2 & -3.7 $\pm$ 0.6 & +0.3 $\pm$ 0.6 \\%
 }{}
  \addlinespace
  \multicolumn{10}{l}{\textbf{Qwen3-8B, penalty $4{\times}10^{-4}$, step 15k (served)}; thresholds $150$k/$190$k, skip $\vpQwenEightKneeSDecode$} \\
  \IfFileExists{generated/qwen8b_serving_rows.tex}{Qwen3-8B $4{\times}10^{-4}$ 15k (served) & $0.75\times Q^*$ & 3 & -10.3 $\pm$ 1.8 & -6.3 $\pm$ 0.8 & -3.4 $\pm$ 0.4 & -2.9 $\pm$ 7.2 & -32.1 $\pm$ 8.4 & -31.5 $\pm$ 5.3 & -0.5 $\pm$ 1.0 \\
Qwen3-8B $4{\times}10^{-4}$ 15k (served) & $0.95\times Q^*$ & 3 & -6.6 $\pm$ 0.3 & -6.0 $\pm$ 0.3 & -3.1 $\pm$ 0.2 & -2.4 $\pm$ 3.5 & -9.1 $\pm$ 0.3 & -8.7 $\pm$ 0.7 & -0.8 $\pm$ 0.3 \\
Qwen3-8B $4{\times}10^{-4}$ 15k (served) & $1.25\times Q^*$ & 3 & -4.9 $\pm$ 0.3 & -4.8 $\pm$ 0.9 & -3.3 $\pm$ 0.4 & -2.0 $\pm$ 0.8 & -5.5 $\pm$ 0.4 & -5.4 $\pm$ 1.1 & -1.0 $\pm$ 1.2 \\%
 }{}
  \addlinespace
  \multicolumn{10}{l}{\textbf{Qwen3-4B, shared Llama thresholds} ($n{=}1$)} \\
  \IfFileExists{generated/qwen_sharedband_rows.tex}{Qwen3-4B (Llama thresholds) & $0.75\times Q^*$ & 1 & +0.9 & +3.6 & +0.9 & -1.0 & +1.8 & +4.0 & +0.0 \\
Qwen3-4B (Llama thresholds) & $0.95\times Q^*$ & 1 & +2.3 & +0.7 & +0.4 & +1.6 & +8.5 & +9.4 & +3.8 \\
Qwen3-4B (Llama thresholds) & $1.25\times Q^*$ & 1 & +1.2 & +1.4 & +0.3 & +1.1 & +1.8 & +2.4 & +3.5 \\%
 }{}
  \bottomrule
  \end{tabular}
\end{table}

\begin{samepage}
\noindent\textbf{Quality and tuning.} Training improves skipping and
accuracy together along the $2{\times}10^{-4}$ run: the later checkpoint skips more and scores higher both natively and when served (Tables~\ref{tab:qwen-serving} and~\ref{tab:qwen-quality}). FlexiDepth's released Llama-3-8B
checkpoint was trained by its authors in two stages, alignment and then
annealing. Our Qwen3 skippers use only the alignment stage, stopped before
one epoch, with a short search over the router penalty, the weight of the
training term that rewards skipping; a larger penalty trades accuracy for
skipping (Appendix~\ref{app:repro}). They are proofs of concept and do not show the best trade-off that full two-stage training can reach on Qwen3. These results motivate further
hyperparameter tuning of the skipper's quality--efficiency trade-off,
which we leave to future work.\par
\end{samepage}

Table~\ref{tab:qwen-quality} repeats the $2{\times}2$ design of
Table~\ref{tab:faithfulness-2x2} for all four checkpoints.
All four difference-in-differences intervals contain zero at one run
per arm, with no resolved loss beyond the checkpoint's own in the
routed comparison.

The separately measured hybrids route
$\vpLqQwenFourHybRoutedShare\%$ (4B) and
$\vpLqQwenEightHybRoutedShare\%$ (8B) of decode rows at the knee and
score $\vpLqQwenFourDelta \pm \vpLqQwenFourDeltaCi$ and
$\vpLqQwenEightDelta \pm \vpLqQwenEightDeltaCi$~pp against always-route.

\begin{table}[h]
  \centering
  \caption{The Qwen3 checkpoints' quality on GSM8K (composite exact
  match, $1{,}319$ held-out documents, bf16): the $2{\times}2$ as in
  Table~\ref{tab:faithfulness-2x2} (left), and the served hybrid at its knee cell, as in Table~\ref{tab:faithfulness-2x2} (right), with routed shares ordered hybrid /
  always-route.}
  \label{tab:qwen-quality}
  \scriptsize
  \setlength{\tabcolsep}{3pt}
  \begin{tabular}{lrrrrrrr}
  \toprule
  Model & A & B & C & D & $(B{-}A)$ & $(D{-}C)$ & DiD (pp) \\
  \midrule
  \IfFileExists{generated/qwen_quality_rows.tex}{Qwen3-4B & 79.45 & 76.80 & 80.14 & 75.74 & -2.65 & -4.40 & -1.74 $\pm$ 1.98 \\
Qwen3-8B $2{\times}10^{-4}$ 10k & 80.97 & 75.44 & 81.12 & 75.28 & -5.53 & -5.84 & -0.30 $\pm$ 1.84 \\
Qwen3-8B $2{\times}10^{-4}$ 18.75k & 80.97 & 77.71 & 81.12 & 78.01 & -3.26 & -3.11 & +0.15 $\pm$ 1.80 \\
Qwen3-8B $4{\times}10^{-4}$ 15k (served) & 80.97 & 70.96 & 81.12 & 70.51 & -10.01 & -10.61 & -0.61 $\pm$ 2.11 \\
 }{}
  \bottomrule
  \end{tabular}
  \vspace{3pt}

  \setlength{\tabcolsep}{2.5pt}
  \begin{tabular}{lrrrrrr}
  \toprule
  Model & upstream & \sys{} (served) & always-route & routed (\%) & vs.\ upstream (pp) & vs.\ always-route (pp) \\
  \midrule
  \IfFileExists{generated/qwen_blockb_rows.tex}{Qwen3-4B & 80.67 & 77.71 & 76.19 & 87 / 100 & $-2.96 \pm 2.15$ & $+1.52 \pm 2.49$ \\
Qwen3-8B $2{\times}10^{-4}$ 10k & 80.36 & 76.65 & 75.21 & 99 / 100 & $-3.71 \pm 2.23$ & $+1.44 \pm 2.60$ \\
Qwen3-8B $2{\times}10^{-4}$ 18.75k & 80.36 & 80.06 & 78.39 & 99 / 100 & $-0.30 \pm 2.27$ & $+1.67 \pm 2.41$ \\
Qwen3-8B $4{\times}10^{-4}$ 15k (served) & 80.36 & 75.89 & 71.65 & 99 / 100 & $-4.47 \pm 2.33$ & $+4.25 \pm 2.68$ \\
 }{}
  \bottomrule
  \end{tabular}
\end{table}

\section{Hardware Transfer}
\label{app:h100}
The protocol of Table~\ref{tab:v13-headline} was repeated on two other
devices with the same code (the A6000 build adds only its decode-graph range and tuned tile table), launch profile, suite, gates and prefill
floor. Each device gets its own knee, its own output lengths, its own decode-graph range, and three paired repetitions against
upstream on that device (Table~\ref{tab:transfer}). The thresholds come from the rule with the bandwidth in Table~\ref{tab:hardware-bands}; no workload-dependent configuration
search was run on any device.

\noindent\textbf{H100 (80\,GB HBM3).} Its upstream $Q^*$ is $\vpLadderHundredGsmKnee$~req/s (Table~\ref{tab:ladders}), with $270$k/$340$k-token thresholds. It uses the A100's tile table unchanged.
End-to-end latency falls on every row with every interval excluding
zero, most at the knee ($\vpHoneEtoEMeanGsmMid\%$). The sign
transfers across the devices, each measured on its own load grid.

\noindent\textbf{RTX A6000 (48\,GB GDDR6, 768\,GB/s).}
By default, upstream captures decode graphs up to $32$ rows on this
card. \sys{} captures up to $256$, the occupancy this card reaches at
its knee, and the upstream control is launched with
\texttt{--cuda-graph-max-bs 256}.
Its $Q^*$ is $\vpLadderAsixGsmKnee$~req/s, with $60$k/$80$k-token thresholds.
The device provides $101$~KB of shared memory per block; the A100's
cohort tiles require $110$~KB.
The tuner selects tiles within the device's shared-memory limit,
holding block-$K$ fixed and requiring bit-identical outputs to the A100 tiles.
It replaced $19$ of $20$ tile keys
(Appendix~\ref{app:kernel-headroom}); the same tuner supports new devices.
Mean end-to-end latency falls on every row, with the interval
excluding zero at the knee and in overload (Table~\ref{tab:transfer}).

\begin{table}[h]
  \centering
  \caption{Bandwidth and compute inputs to the rule, estimated crossovers
  and served decode thresholds.
  Estimates use $V^* = \tau\,\mathrm{BW} / (s L_r b)$ with
  $\tau = \vpBandTauMs$~ms, $s = \vpBandSkipRatio$, $L_r = 16$ and $b = 4{,}096$ bytes.}
  \label{tab:hardware-bands}
  \scriptsize
\begin{tabular}{lrrrrr}
\toprule
Device & BW (GB/s) & TFLOPS & $V^*$ & exit / enter & $Q^*$ (req/s) \\
\midrule
\IfFileExists{generated/hardware_bands_rows.tex}{A100-SXM4-80GB & 1,935 & 312 & 157.7k & 160k / 200k & 13 \\
H100 80\,GB HBM3 & 3,350 & 990 & 273.0k & 270k / 340k & 14 \\
RTX A6000 48\,GB GDDR6 & 768 & 155 & 62.6k & 60k / 80k & 6 \\
 }{}
\bottomrule
\end{tabular}
\end{table}

\begin{table}[h]
  \centering
  \caption{The upstream $Q^*$ sweeps: contiguous integer offered
  rates, $Q^*$, achieved output tok/s at $Q^*$ and at the
  highest rate, and the growth left above $Q^*$ (computed from unrounded
  values).}
  \label{tab:ladders}
  \scriptsize
  \begin{tabular}{llrrrr}
  \toprule
  Workload & rates (req/s) & $Q^*$ & tok/s at $Q^*$ & tok/s at top & growth above $Q^*$ (\%) \\
  \midrule
  \IfFileExists{generated/ladder_rows.tex}{GSM8K & 8--18 & 13 & 1079 & 1080 & +0.1 \\
BBH & 20--36 & 34 & 3121 & 3106 & -0.5 \\
CoQA & 16--33 & 25 & 1216 & 1217 & +0.0 \\
 }{}
  \IfFileExists{generated/ladder_h100_rows.tex}{GSM8K (H100) & 12--22 & 14 & 1226 & 1228 & +0.1 \\
 }{}
  \IfFileExists{generated/ladder_a6000_rows.tex}{GSM8K (RTX A6000) & 2--10 & 6 & 493 & 492 & -0.2 \\
 }{}
  \bottomrule
  \end{tabular}
\end{table}

\section{Reproducibility}
\label{app:repro-dtype-note}
\label{app:repro}

\noindent\textbf{Models.} Base: For Llama-3-8B, \texttt{NousResearch/Meta-Llama-3-8B-Instruct}
at revision \texttt{53346005fb0e}. Skipper: the released
\texttt{xuan-luo/FlexiDepth-Llama-3-8B-Instruct} (routes the last 16 of
32 layers),
served as published. Qwen3-4B: \texttt{Qwen/Qwen3-4B} at
\texttt{1cfa9a72}; Qwen3-8B: \texttt{Qwen/Qwen3-8B} at \texttt{d117af2f};
both with thinking disabled through the chat template. Their skippers
are released with the artifact: the served checkpoints as full exports,
the two other Qwen3-8B checkpoints as router deltas, each with its probe
and gate records.

\noindent\textbf{Training of the Qwen3 skippers.}
Both use alignment-only FlexiDepth training: the router and projector
are trained over a frozen base with a straight-through hard gate.
Training uses \texttt{allenai/tulu-3-sft-mixture}, sequence length
$2{,}048$, lr $10^{-4}$, global batch $32$ and at most one epoch
($27{,}643$ steps).
Checkpoints are saved every $1{,}250$ steps with a 100-document scoring probe and a 32-prompt routing probe that reports the chat-template skip share.

Training varies the router penalty using the probe's skip share
and score.
The effective coefficient is penalty divided by accumulation, with
accumulation fixed at $4$.
Qwen3-4B uses $10^{-5}/4$ through step $20{,}000$.
Qwen3-8B uses $10^{-4}/4$ for $7{,}500$ steps, then
$2\times10^{-4}/4$ or $4\times10^{-4}/4$ to step $20{,}000$.
The routing probe's skip shares of the reported checkpoints are
$\vpProbeSkipQwenEightOld$ and $\vpProbeSkipQwenEightAlt$ at steps
$10{,}000$ and $18{,}750$ of the $2\times10^{-4}$ run and
$\vpProbeSkipQwenEight$ at step $15{,}000$ of the $4\times10^{-4}$ run,
which is served.

Both ran on a CloudLab node with four A100-SXM4-40GB GPUs.
The trainer and export scripts are in
\texttt{vskipper/training/qwen3-flexidepth/}, alongside the input files and gate records.

\noindent\textbf{Engine.} Baseline: upstream SGLang at commit
\texttt{602c8615a1}, a separate checkout.
\sys{}: the same commit plus the runtime.
Both use the shared launch profile and graph-capture settings
of \S\ref{sec:methodology}.
Appendix~\ref{app:gates} specifies the boot-configuration and execution-difference checks.
The switch's constants live in one module and are attested in every cell.

\noindent\textbf{$Q^*$ sweeps.} Table~\ref{tab:ladders} gives the upstream
rate sweeps used to measure $Q^*$ (\S\ref{sec:methodology}).
A rate counts as growth when it and the next rate both exceed the
running maximum by at least $1\%$.
The Qwen $Q^*$ values use the same rule.

\noindent\textbf{Cost of a rerun.}
The measured cells sum to about $34$ GPU-hours on the A100: main cells $12.5$, ablations $5.2$, RandomSkip $7.4$ and Qwen $8.9$.
The H100 cells take $1.8$ GPU-hours and the RTX A6000 cells $4.2$.
Rate sweeps, length-recording runs, quality cells and restarts roughly double the
A100 figure.
At the rental prices in Table~\ref{tab:instances}, the serving
evidence reruns for under USD~$100$; the A100 price covers four GPUs.
The Qwen3-4B skipper trains in about $40$ A100-40GB GPU-hours; each
Qwen3-8B run, trained to step $20{,}000$, in about $60$.

\begin{table}[t]
  \centering
  \caption{The rented Vast.ai machines behind the serving cells, by host
  id (instance numbers change on every rental); driver as reported by
  the node, prices as rented, DLPerf as Vast.ai reports it for the
  whole machine.}
  \label{tab:instances}
  \scriptsize
  \setlength{\tabcolsep}{3pt}
  \begin{tabular}{lp{3.6cm}p{3.9cm}llrl}
  \toprule
  Host & CPU & GPU & Price & Driver & DLPerf & Location \\
  \midrule
  458442 & $2\times$ AMD EPYC 7713 & $4\times$ A100-SXM4-80GB (NVLink) & \$4.60/h & 595.71 & 469.1 & Japan \\
  638511 & AMD EPYC 9454 (24 vCPU) & $1\times$ H100 80\,GB HBM3 (SXM) & \$2.09/h & 595.71 & 314.6 & Germany \\
  314882 & $2\times$ Intel Xeon E5-2699 v4 & $1\times$ RTX A6000 48\,GB & \$0.49/h & 580.173 & 54.5 & Virginia, US \\
  \bottomrule
  \end{tabular}
\end{table}

\end{document}